# Bootstrapped Dimensional Crossover of a Spin Density Wave

Anjana M. Samarakoon[1], J. Strempfer[2], Junjie Zhang[1,3], Feng Ye[4], Yiming Qiu[5], J.-W. Kim[2], H. Zheng[1], S. Rosenkranz[1], M. R. Norman[1], J. F. Mitchell[1] and D. Phelan[1]

[1]*Materials Science Division, Argonne National Laboratory, Lemont, IL 60439, USA*

[2]*Advanced Photon Source, Argonne National Laboratory, Lemont, IL 60439, USA*

[3]*State Key Laboratory of Crystal Materials, Shandong University, 250100, Jinan, Shandong, China*

[4]*Neutron Scattering Division, Oak Ridge National Laboratory, Oak Ridge, TN 37830, USA*

[5]*NIST Center for Neutron Research, National Institute of Standards and Technology, Gaithersburg, MD 20899, USA*

## Abstract

Quantum materials display rich and myriad types of magnetic, electronic, and structural ordering, often with these ordering modes either competing with one another or 'intertwining,' that is, reinforcing one another. Low dimensional quantum materials, influenced strongly by competing interactions and/or geometric frustration, are particularly susceptible to such ordering phenomena and thus offer fertile ground for understanding the consequent emergent collective quantum phenomena. Such is the case of the quasi-2D materials $R_4Ni_3O_{10}$ (R=La, Pr), in which intertwined charge- and spin-density waves (CDW and SDW) on the Ni sublattice have been identified and characterized. Not unexpectedly, these density waves are largely 2D as a result of weak coupling between planes, compounded with magnetic frustration. In the case of R=Pr, however, we show here that exchange coupling between the transition metal and rare earth sublattices upon cooling overcomes both obstacles, leading to a dimensional crossover into a fully 3D ordered and coupled SDW state on both sublattices, as an induced moment on notionally nonmagnetic $Pr^{3+}$ opens exchange pathways in the third dimension. In the process, the structure of the SDW on the Ni sublattice is irreversibly altered, an effect that survives reheating of the material until the underlying CDW melts. This 'bootstrapping' mechanism linking incommensurate SDWs on the two sublattices illustrates a new member of the multitude of quantum states that low-dimensional magnets can express, driven by coupled orders and modulated by frustrated exchange pathways.

# Introduction

Understanding and manipulating intertwined order parameters associated with density waves is widely recognized as a frontier challenge in quantum materials [1,2], with relevance to our understanding of, for example, high temperature superconducting cuprates [3–5], twisted bilayer graphene [6,7], and superfluid $^4$He [8,9]. Even in more conventional metals with itinerant *d* electrons, the electron-phonon interaction can lead to charge density waves (CDW) intertwined with antiferromagnetic spin density waves (SDW) whose wavelengths are related as $\lambda_{SDW} = 2\lambda_{CDW}$ [10–12]. Importantly, the character of density waves is often influenced by reduced structural dimension, where the 1D- or 2D-nature of the crystal structure can set the dimensionality of the density waves. However, this need not be the case, as shown by materials such as $CuIr_2S_4$ [13], in which the electronic dimensionality (1D), and hence the density wave order, is lower than the structural dimensionality (3D). Alternatively, weak interchain coupling seen, for example, in organic salts, can lead to density wave dimensionality higher than that of the underlying crystal structure [14]. Such behaviors demonstrate that the connection between crystal structure and density wave dimensionality can be malleable. To this point, strong magnetic fields can induce 1D- to 3D SDW crossovers in Bechgaard salts such as $(TMTSF)_2ClO_4$ [15], and a 2D- to 3D-dimensional crossover is found in the CDW phase of layered $La_{2-x}Ba_xCuO_4$ upon cooling below ≈40 K [16].

Here, we introduce another layered transition metal oxide, $Pr_4Ni_3O_{10}$, in this case with two magnetic sublattices, in which a quasi-2D incommensurate SDW structure, established at a metal-to-metal transition $T_{MMT} \approx 158$ K, transforms with cooling into a long range ordered 3D structure of stacked SDWs at $T_f \approx 26$ K. In this material, a singular form of magnetic dimensional crossover occurs through a "bootstrapping" mechanism, whereby the SDW of the 2D-ordered transition metal sublattice induces a counterpart SDW on the rare earth sublattice, which would otherwise be non-magnetic due to the singlet crystal field ground state of the non-Kramers $Pr^{3+}$ ion in a low symmetry site. In turn, the induced order of the second sublattice opens a new magnetic exchange pathway perpendicular to the layers, allowing it to bootstrap the 3D ordered state of *both* rare earth and transition metal sublattices. The resulting inter-layer correlations, expressed through a reconstructed transition metal SDW, become frozen in as a metastable magnetic configuration. Upon reheating even well above the rare earth ordering temperature $T_f$, these correlations survive on the Ni sublattice, albeit with a greatly reduced 3D correlation length, until they are finally erased by the melting of the intertwined density wave state at $T_{MMT}$. We propose a model spin Hamiltonian to explain these phenomena, including the evolution of magnetic states, the thermal hysteresis, and the observed 2D to 3D cross-over.

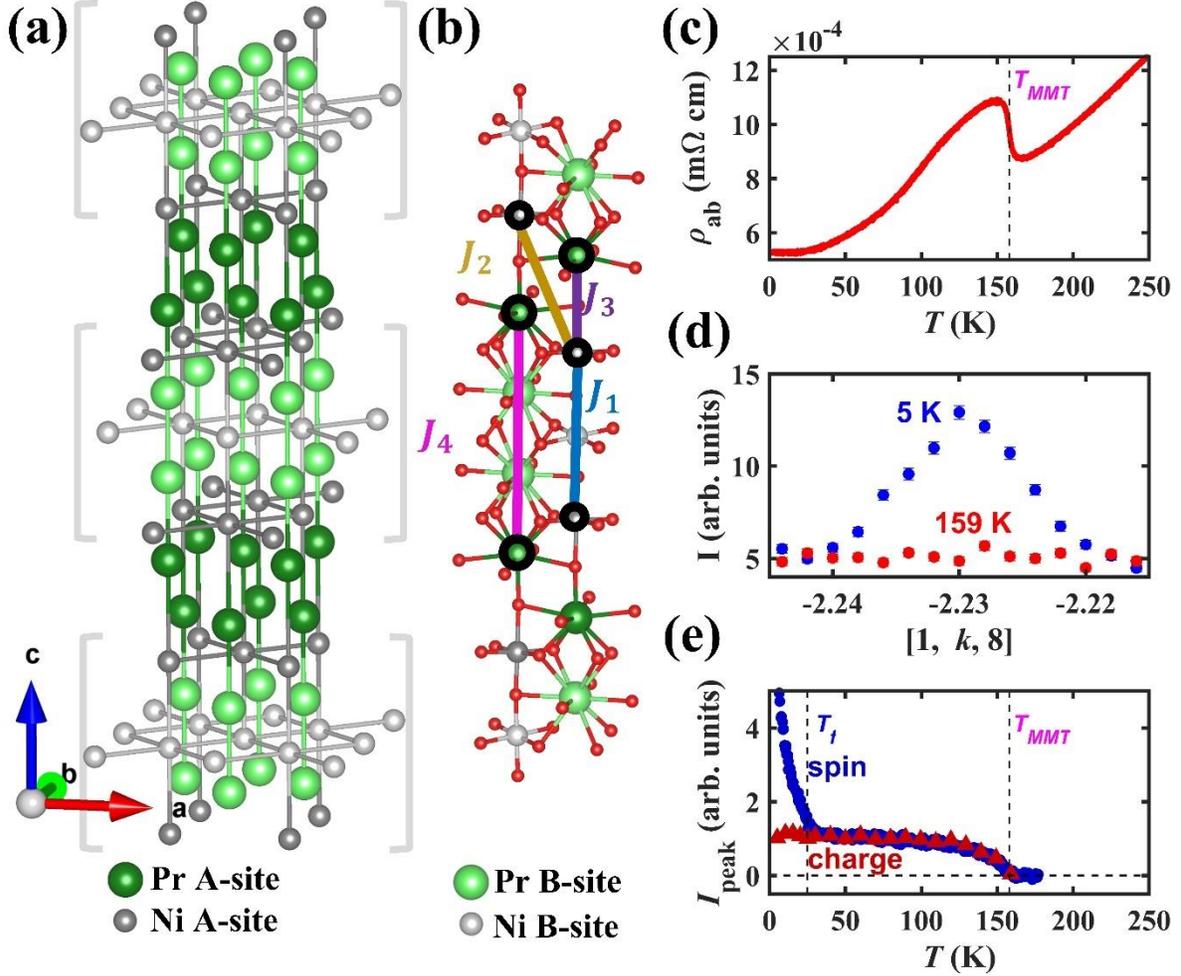

Figure 1: (a) Crystal structure of $R_4Ni_3O_{10}$ ($R$ = La, Pr), with oxygen ions omitted for clarity. The brackets indicate the nickel trilayers. (b) A part of the nuclear unit cell with corresponding O ligands. The magnetic exchange interactions of the effective spin-Hamiltonian discussed in the text are also marked as single-color solid lines with open-circle ends, and the color scheme matches the labels of the exchanges ($J_x$, $x = $ 1,2,3,4). Here, $J_1$ and $J_2$ couple Ni A sites, $J_3$ couples Ni A sites with Pr A sites, and $J_4$ couples Pr A sites. (c) In-plane resistivity ($\rho_{ab}$) of $Pr_4Ni_3O_{10}$ [17]. (d) CDW peak measured with X-rays along [1,$k$,8] at 5 K and the background at 159 K. (e) Normalized temperature dependence of the CDW peak measured by X-rays at $\bm{Q}$=(1,-2.23,8) (red triangles) and the SDW peak at $\bm{Q}$ = (0,-0.6,-4.6), which lies close to one of the maxima in the low-temperature phase (blue circles). Note that the breadth of the peak and limited wave-vector resolution along $\bm{c}^*$ allowed for the capture of the order parameter despite a 0.1 r.l.u. offset from the peak center.

The crystal structure of $R_4Ni_3O_{10}$ ($R$ = Pr, La) is formed by nickel oxide trilayer slabs ($R_2Ni_3O_8$) separated by rare earth oxide bilayer slabs ($R_2O_2$). As shown in Fig. 1a, the two sublattices are interleaved by a lattice centering translation, in which Ni trilayers and $R$ bilayers alternate (see Fig. 1b and Suppl. Fig. S14). As discussed in Ref. [17], the space group of $R_4Ni_3O_{10}$ is monoclinic, $P2_1/a$. However, the deviation from orthorhombic $Bmab$ is small, and we use this symmetry herein. The $B$-centering of the lattice leads to an in-plane translation of (1/2,0) between successive Ni trilayer blocks along $\bm{c}$. Thus, within a Ni trilayer, the Ni layers are in registry (since they are separated by one $R_2O_2$ layer), but subsequent trilayers are out of registry (since they are separated by two $R_2O_2$ layers). This atomic arrangement is expected to favor 2D antiferromagnetic order within the trilayer blocks yet simultaneously hinder a truly long-range ordered 3D magnetic state of the Ni $3d$ electrons because of (*i*) a substantial spatial separation along the ***c***-axis through the $R_2O_2$ block and (*ii*) a geometric frustration akin to that found in $K_2NiF_4$ [18] that hinders the development of a unique 3D ground state for antiferromagnetic interactions of ordered moments between neighboring trilayers.

Both $La_4Ni_3O_{10}$ and $Pr_4Ni_3O_{10}$ undergo metal-metal transitions that are manifest as anomalies in electrical transport data at $T_{MMT}$ [19,20], which is ≈148 K for $La_4Ni_3O_{10}$ and ≈158 K for $Pr_4Ni_3O_{10}$ [17], as shown in Fig. 1c. Recent single-crystal X-ray and neutron diffraction work on $La_4Ni_3O_{10}$ established that incommensurate magnetic and structural peaks appear at $T_{MMT}$ upon cooling [12]. Augmented by band-structure calculations, these observations were interpreted as a coupled CDW and antiferromagnetic SDW with twice the charge wavelength in a Fermi-surface nesting driven picture. Importantly, all charge and spin ordering in $La_4Ni_3O_{10}$ is thought to occur within the trilayer perovskite units, and specifically, the CDW and SDW are centered on nickel sites based on structure factor calculations [12]. We proposed a model in which the SDWs with opposite polarization occur on the upper and lower layers within a Ni trilayer with the middle Ni-layer being non-magnetic. The core electrons of $La^{3+}$ cations are uninvolved in the CDW, and given that it is non-magnetic, the role of $La^{3+}$ is entirely steric. $Pr^{3+}$ cations, on the other hand, have partially filled *f*-shells ($4f^2$), which makes magnetic ordering possible for this sublattice, though this also requires an appreciable exchange field to overcome its symmetry governed crystal field singlet ground-state (this is discussed in detail in the Supplement Section VI). It also opens the potential for inter-sublattice $3d$-$4f$ exchange interactions in $Pr_4Ni_3O_{10}$. Indeed, here we show that such exchange pathways provide a mechanism for overcoming *both* roadblocks to 3D order discussed above—inter-trilayer separation and geometric frustration—leading to the novel bootstrapping and imprinting behavior expressed by this intertwined density wave system.

## Results

### Experiments

Consistent with the convention of Ref. [12], in diffraction experiments we refer to a face-centered unit cell of dimensions ≈ 5.4 × 5.4 × 27 Å$^3$, such that the CDW wavevector is along ***b***, which is at a 45° angle to the Ni-O bonds in the basal plane (see Fig. 1a). This setting is a $\sqrt{2} \times \sqrt{2} \times 1$ supercell of a body centered tetragonal setting with dimensions ≈ 3.9 × 3.9 × 27 Å$^3$. Unpolarized X-ray diffraction measurements, sensitive to charge, were taken through the anticipated CDW wave-vector (see Ref. [12]), and a peak was only observed below $T_{MMT}$ as shown in Fig. 1d at ***Q*** = (1, -2.23, 8), with an order parameter labeled as "charge" in Fig. 1e. This charge order parameter turns on at $T_{MMT}$. Measurements of a CDW peak performed as a function of X-ray energy near the Ni *K*-edge showed a shift in the energy at which the CDW peak was observed compared to the fluorescence (see Fig. 2a). This is a hallmark of a contribution from the anomalous Ni scattering and is thus indicative of a significant contribution of the nickel electrons to the CDW.

Thus, in concert with our previous band structure calculations and structure factor arguments [12], we confirm that the CDW in $Pr_4Ni_3O_{10}$ originates from Ni $3d$ electrons. We note that the CDW order parameter (Fig. 1e) shows no anomalies below $T_{MMT}$; rather, the temperature dependence is consistent with a single transition at $T_{MMT}$. We conclude from these observations that $La_4Ni_3O_{10}$ and $Pr_4Ni_3O_{10}$ behave alike in the charge sector.

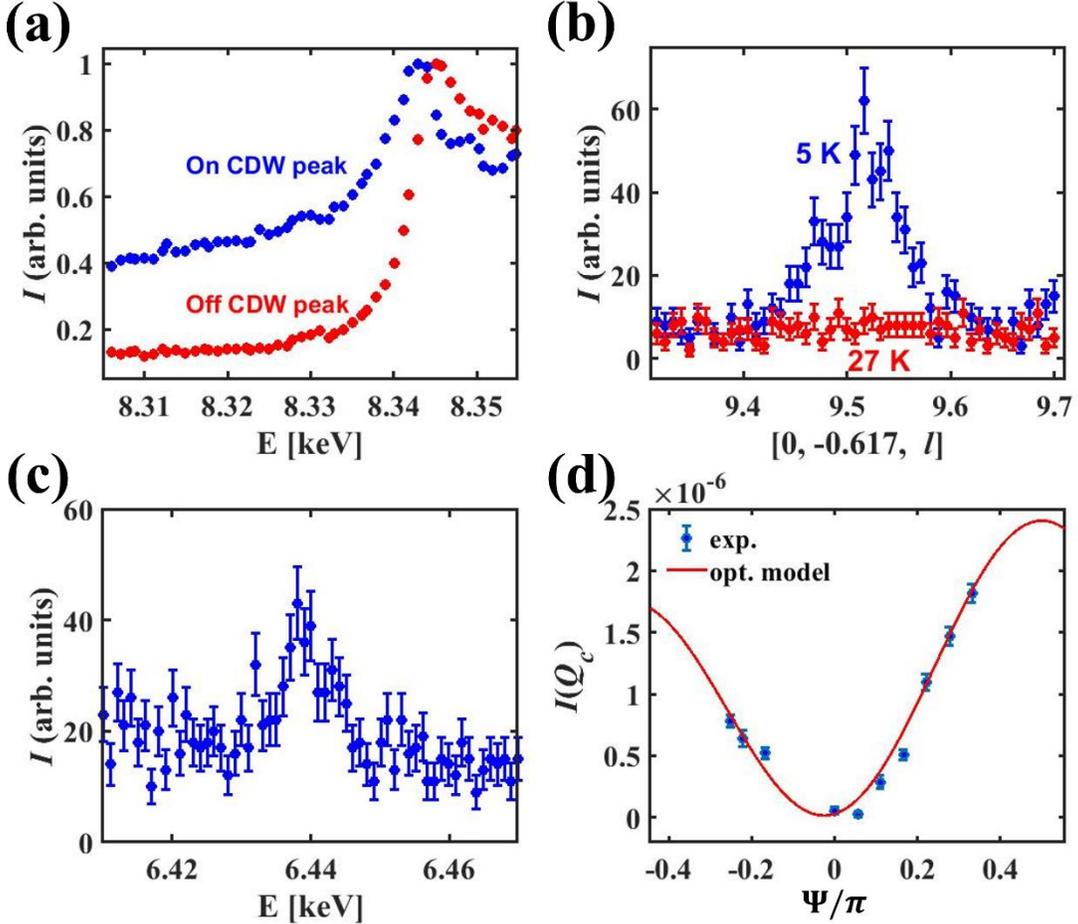

Figure 2: Resonant X-Ray scattering. (a) Energy scans around the Ni K edge on and off the CDW Bragg peak showing the energy dependence of the CDW peak and the fluorescence background, respectively. (b) The polarized resonant peak at $\boldsymbol{Q} = (0, -0.617, 9.5)$ showing the Pr magnetic moment ordering at 5 K but not at 27 K. (c) Energy dependence of the peak showing resonance at the Pr $L_2$ edge. (d) The resonant peak integrated intensity as a function of the azimuthal angle $\Psi$ in comparison to (solid red line) that calculated from the optimized model with Pr moments along $\boldsymbol{b}$. The resonant peak itself as a function of $\Psi$ is shown in Suppl. Fig. S7a. Note that all error bars in this and other figures represent one standard deviation.

Figure 3 (a and b) shows the neutron diffraction intensities collected in the (0$kl$) scattering plane for both $Pr_4Ni_3O_{10}$ (left panels) and $La_4Ni_3O_{10}$ (right panels) at 6 K and 50 K. The CDW peaks at the wave-vectors anticipated from the X-ray experiments are apparently too weak to be observed due to the significantly lower flux in the neutron experiments compared to the synchrotron x-rays, and the temperature-dependent superlattice peaks with $\boldsymbol{q}_{SDW} \approx (0, 0.6, 0)$ are magnetic. Note that corresponding measurements at $T = 200$ K ($> T_{MMT}$) have been treated as a non-magnetic background and subtracted from the rest of the data. At 50 K, which is below $T_{MMT}$ for both samples, similar cigar-shaped magnetic peaks are observed at $k \approx 0.6$ and 1.4 and $l \approx 2$ and 6. Incommensurability in $k$ can be understood as an in-plane SDW coupled to the CDW, obeying the relation $2q_{SDW} = q_{CDW}$ [12]. The readily apparent anisotropic nature of the magnetic peaks clearly indicates that the magnetic order is quasi-2D within the basal plane. However, as shown in Fig. 3 (c and d), the $l$-dependence of each incommensurate peak is complex. As discussed below, extensive modeling of short-ranged order among multiple layer stacking patterns is required to describe this scattering intensity, including both $\boldsymbol{Q}$-dependences and peak widths. In contrast, CDW peaks as shown in Fig. 2b are significantly narrower, indicative of longer-range correlations along $\boldsymbol{c}$ in the charge sector.

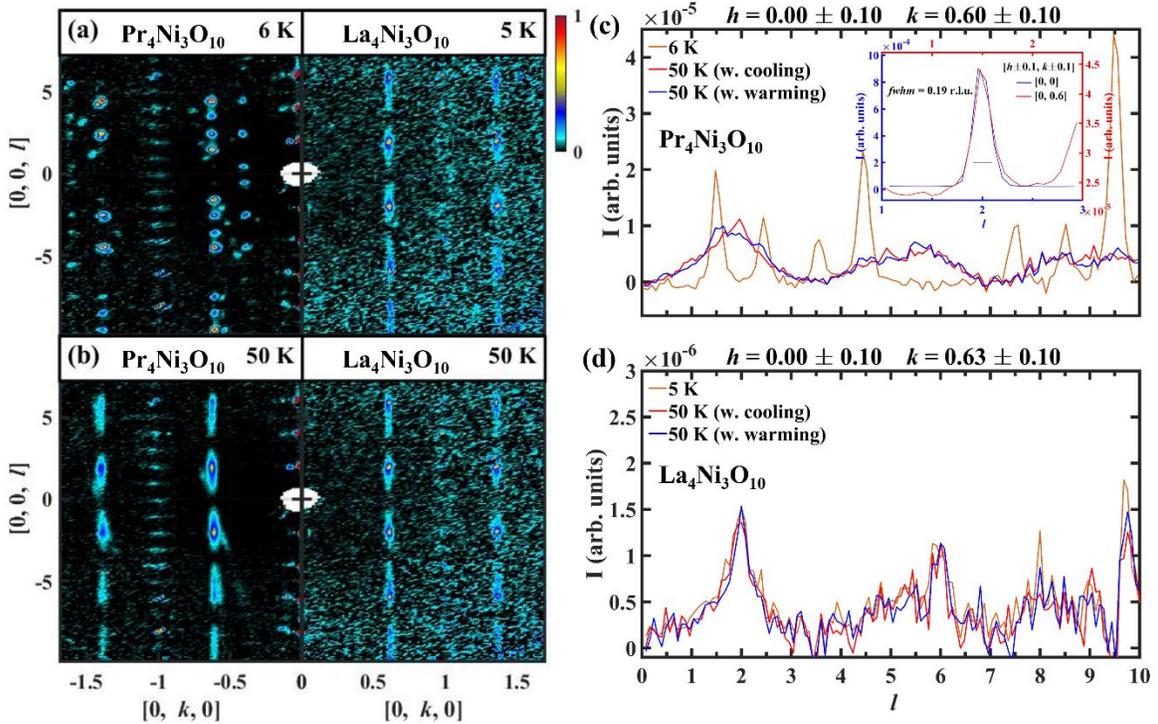

Figure 3: Neutron scattering data collected on CORELLI. Slices along $[0kl]$ for (left) Pr$_4$Ni$_3$O$_{10}$ and (right) La$_4$Ni$_3$O$_{10}$. Samples measured at (a) $\approx 6$ K and (b) 50 K after cooling directly from the paramagnetic phase. Panels (c) and (d) show the line cuts along $l$ at $h = 0\pm 0.1$ and $k =0.6\pm 0.1$ for Pr$_4$Ni$_3$O$_{10}$ and La$_4$Ni$_3$O$_{10}$, respectively. Data were collected at 50 K twice while cooling to and warming from base temperature ($\approx 6$ K). The data measured at 200 K ($> T_{MMT}$) were treated as a non-magnetic background and subtracted (see Suppl. Section I). Inset: The nuclear Bragg peaks along $l$ at $h= 0\pm 0.1$ and $k=0\pm 0.1$ for Pr$_4$Ni$_3$O$_{10}$ are compared to the line cut through the SDW. The similar line-shape indicates a lack of distinguishable finite size broadening along $\boldsymbol{c}$ within the constraint of the wave-vector resolution of the measurement on CORELLI. The in-plane correlation length of the SDW below $T_{MMT}$ is also comparable to that of the CDW and the nuclear structure.

The neutron diffraction pattern for La$_4$Ni$_3$O$_{10}$ remains unchanged at 6 K compared to 50 K as it contains only a single magnetic phase transition at $T_{MMT}$ due to the Ni 3$d$ electrons. In contrast, a quite different pattern is observed at 6 K for Pr$_4$Ni$_3$O$_{10}$. The magnetic peaks appear at the same incommensurate values of $k$ as at 50 K, but the $l$-dependence, intensities, and shapes of the peaks are all dramatically modified. The peaks now appear sharp at $l = n/2$ (where $n$ is odd), which indicates that the magnetic unit cell has doubled along **c**. As shown below, this reflects an irreversible alteration of the SDW on the Ni sublattice as well as an ordering of Pr moments. The breadth of the peaks along $l$ has drastically diminished (see Fig. 3c), reflecting a cross-over to 3D magnetic order (see inset, Fig. 3c). Additionally, the significantly enhanced intensity indicates an increase in the total ordered moment. This observation, combined with the absence of such phenomena in La$_4$Ni$_3$O$_{10}$, suggests that the Pr 4$f$ electrons are responsible for driving the system from quasi-2D order to 3D order. The order parameter associated with the two magnetic transitions can be tracked by the temperature-dependent intensity at fixed $\mathbf{Q} = (0, -0.6, -4.6)$, which overlaps with the magnetic intensities in both the lower $T$ and higher $T$ Pr$_4$Ni$_3$O$_{10}$ phases and is displayed as the "spin" order parameter in Fig. 1e. The gradual change with temperature below $T_f \approx 26$ K, indicative of the second ordering, is suggestive of an induced moment ordering, presumably on Pr, rather than a conventional phase transition [21–25].

To confirm this Pr-specific moment ordering, resonant polarized X-ray scattering experiments were performed around the Pr $L_2$ edge (see *Methods: X-ray resonant scattering*). Figure 2b shows scans through $\mathbf{Q}_{RXS} = (0, -0.617, 9.5)$. Scattering in the σ–π′ polarization channel renders this configuration sensitive only to magnetic order, and the resonance makes the sensitivity specific to Pr. Indeed, a temperature-dependent peak, visible at 5 K but not at 27 K, confirms the ordering of Pr 4$f$ moments in the ground-state phase, as the X-ray energy dependence of the peak shown in Fig. 2c is consistent with resonance at the Pr $L_2$ edge. Furthermore, we observed a gradual increase in the strength of this peak below 27 K (see Suppl. Fig. S7b), and the temperature trend again is consistent with an induced moment transition. Finally, we observed an azimuthal ($\Psi$) dependence of the resonant scattering, as shown in Fig. 2d, that indicates the direction of the Pr moment is approximately along **b** at 5 K (more information can be found in Suppl. Section II).

An examination of the $T$-dependent neutron diffraction intensity along $l$ allows us to understand qualitatively the relationship of the spin order between the low $T$-state ($T < T_f$), the intermediate $T$ state ($T_f < T < T_{MMT}$) and the high $T$ state ($T > T_{MMT}$) of Pr$_4$Ni$_3$O$_{10}$ (see Suppl. Sect. I). As shown in Fig. 4a and 4c, the line shapes in the intermediate temperature state differ markedly on cooling versus warming. Specifically, a comparison of the diffuse signal in this intermediate temperature state reveals a clear spectral weight distribution biased towards $l \approx \frac{n}{2}$ ($n$ odd) during warming from the low $T$ state. Thus, the **c**-axis ordering of the SDW established in the low-$T$ structure with ordered Pr moments *imprints* itself onto the Ni magnetic sublattice on warming, even in the absence of rare earth magnetic order. For $T_f < T < T_{MMT}$ the imprinting manifests as a short-range order that was absent on cooling. This kind of metastable imprinting on the SDW configuration, and the concomitant hysteresis in the scattering pattern, is presumably driven by the Ni-3$d$ to Pr-4$f$ exchange coupling. Noting that this hysteretic imprinting on the Ni-centered SDW and the accompanying modification of its magnetic structure have not been observed previously to our knowledge, we now provide quantitative models that justify these claims.

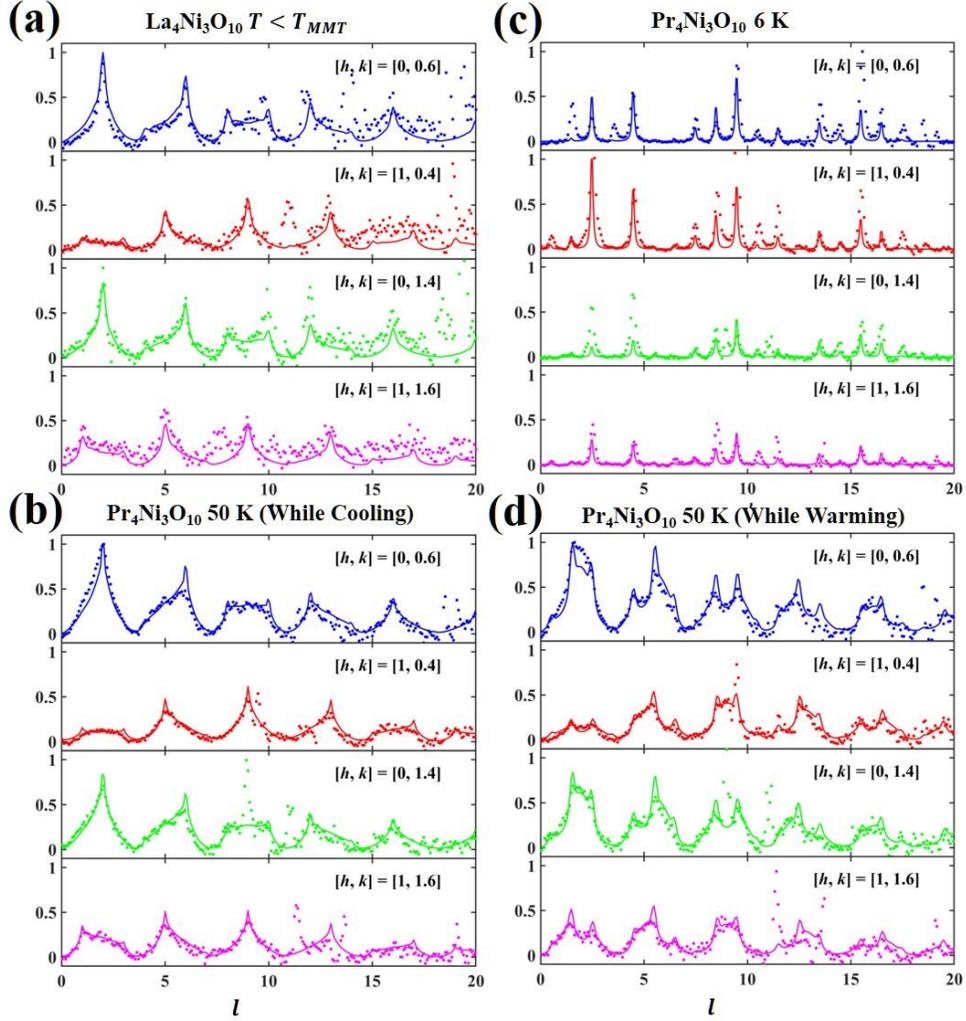

Figure 4: Comparison of neutron data (symbols) and the corresponding optimized magnetic models (solid curves). (a) $La_4Ni_3O_{10}$ data collected at $T < T_{MMT}$, (b) $Pr_4Ni_3O_{10}$ data at 50 K measured while cooling from 200 K, (c) $Pr_4Ni_3O_{10}$ data collected at base-temperature and (d) $Pr_4Ni_3O_{10}$ data measured at 50 K after warming from base-temperature ($\approx$ 6 K). Because of the temperature independent nature of the structure factors below $T_{MMT}$ for $La_4Ni_3O_{10}$ (see Fig. 3d), all the low temperature data (5 K, 50 K warming and cooling) were combined in panel (a). Each panel shows line cuts of 3D scattering data along $l$ for four different combinations of $h \pm 0.1$ and $k \pm 0.1$: [0, 0.6], [1, 0.4], [0, 1.4] and [1, 1.6]. The data measured at 200 K ($> T_{MMT}$) were treated as the non-magnetic background and subtracted (see Suppl. Section I).

**Modeling and data fitting**

We explored a variety of numerical models accompanied by an iterative optimization procedure (IOP) to fit the experimentally measured neutron scattering intensities, $I^{exp}(\boldsymbol{Q})$ (see *Methods: Optimization*) for the ordered state in La$_4$Ni$_3$O$_{10}$, as well as the three proposed ordered states of Pr$_4$Ni$_3$O$_{10}$: the ground-state ($LowT$), the intermediate temperature state on warming ($IntW$), and the intermediate temperature state on cooling ($IntC$).

We begin by considering the case of La$_4$Ni$_3$O$_{10}$, which exhibits only a single, 2D-like magnetic phase. Based on limited data collected with a triple-axis spectrometer along *l*, we previously proposed [12] that each individual Ni trilayer consists of a simple sinusoidal SDW in the top and bottom layers of equal magnitude with a π phase difference between the two layers; the middle layer has no ordered moment in this picture. As alluded to above, next neighboring trilayers are structurally equivalent under a translational symmetry operation of [½, 0, ½] (equivalent to a body-centering translation in the tetragonal setting), and the proposed SDW phase is invariant under this operation. The sequence of stacking of the six individual Ni layers which compose a magnetic unit cell can thus be expressed as a series of multiplicative factors denoted by [1,0,-1], meaning the top and bottom layers of a trilayer are π out of phase (i.e., antiferromagnetically aligned) and the middle layer lies on a node. This 2D structure is illustrated in Fig. 5a.

The more complete data for La$_4$Ni$_3$O$_{10}$ collected on CORELLI now allows for a more thorough determination of the correlations. First, we find that the direction of spin polarization in the top and bottom layers lies along $\boldsymbol{a}$, perpendicular to the direction of propagation of the SDW, $\boldsymbol{q}_{SDW} \perp \boldsymbol{a}$. Symmetry demands that the ordered moments of the middle layer vanish, unless the arrangement of spins is noncollinear. We have tested for this latter possibility and find that, within the sensitivity of the data, there is an upper bound on any noncollinear ordered moment in the middle layer of ≈ 60% that of the ordered moment in the outer planes ($M_{Ni_B} < 0.6 \cdot M_{Ni_A}$). For convenience, we have set the ordered moment of the middle layer to be zero in the rest of the numerical analysis.

Given this framework, we then consider a model (see Suppl. Section III) consisting of a superposition of correlated trilayers of varying stacking length $L_n$, where $L_n$ refers to the number of trilayers that are correlated along $\boldsymbol{c}$ (each possessing the series of multiplicative phase factors [1,0,-1]); that is, $L_1$ refers to a single trilayer uncorrelated with its neighbors, whereas $L_2$ refers to two (and only two) correlated trilayers. Thus, the intensity can be considered to be proportional to $\sum_n \rho(L_n) \left[\frac{I(L_n)}{L_n}\right]$, where $\rho(L_n)$ represents the probability that a given layer lies within a correlated block of size $n$ trilayers and $I(L_n)$ represents the corresponding structure factor [note that $I(L_n)$ is proportional to $L_n$]. The fitting of $\rho(L_n)$ allowed for optimization of the observed peak shapes, which were complex and contained both broad (captured by low *n*) and narrow components (captured by high *n*). The best fit, as shown in Fig. 4a, consisted of a rapidly decaying $\rho(L_n)$ with increasing *n* – as would be expected for quasi-2D order – combined with a nonzero minority contribution captured by a component with larger $n$ which breaks the trend of decay. This model implies what appears to be an inhomogeneous mixture in which the majority of the SDW order is short-ranged (with low $n$) but for which a minority long-range ordered component (with high $n$) coexists. Let us note that this latter minority component can be captured by any sufficiently high value of $n$ that approaches the resolution limit of the spectrometer and so the exact value of $n$ should not be considered significant. Further details of the peak fitting are described in Figs. S11 and S12 and surrounding text. Application of the same approach for the $IntC$ state of $Pr_4Ni_3O_{10}$ also leads to a reasonable fit of the observed $l$-dependence (see Fig. 4b), and a rapidly decaying $\rho(L_n)$, with a similar cut-off ($n > 3$ in this case), consistent with the pseudo-2D structure and polarization along $\boldsymbol{a}$ found for $La_4Ni_3O_{10}$. The outcome of this analysis has confirmed the qualitative picture developed in Ref. [12] for this intermediate temperature structure, which is shared between La- and $Pr_4Ni_3O_{10}$ (on cooling), but has established a more quantitative foundation for how the trilayer building blocks correlate with their neighbors.

We now turn to the low temperature behavior of $Pr_4Ni_3O_{10}$. To construct a SDW model for the $LowT$ phase, we considered a magnetic unit-cell of twice the nuclear unit-cell along $\boldsymbol{c}$ with periodic boundary conditions since the magnetic correlations clearly span many nuclear cells along $\boldsymbol{c}$, and Bragg peaks are located at half-integer positions (Fig. 3c). In the subsequent section, we construct a phenomenological model to explain this cell doubling. As shown above, resonant X-ray scattering unambiguously shows that Pr contributes to the magnetic scattering. Thus, an SDW model for the low temperature phase must explicitly account for this contribution. Indeed, attempts to fit the low temperature data using a Ni-only model fail (see Suppl. Section IV, Fig. S13). Note that as shown in Fig. 1a, the basal planes of Pr atoms can be subdivided into two inequivalent sites marked as A and B. The planes consisting of Pr A sites lie between Ni trilayers while the planes consisting of Pr B sites lie within the Ni trilayers. Since we see strong magnetic peaks along $\boldsymbol{c}$ with the same in-plane incommensurate wavevectors $(h, k)$ as found above the dimensional crossover, the Pr-layers are treated as an induced SDW, with the same in-plane periodicity as the Ni-layers. However, the SDW polarization (moment orientation) $\boldsymbol{\sigma}_{Pr}$ for each Pr-layer, the SDW polarization $\boldsymbol{\sigma}_{Ni}$ for Ni-layers, and the amplitude of SDWs for Pr A and B sites ($M_{Pr_A}$, $M_{Pr_B}$) were refined independently using IOP. Furthermore, not only was the neutron structure factor considered, but also $I(\boldsymbol{Q}_{RXS}, \Psi)$ from the RXS experiment was included in the optimization. This optimization process concludes that the induced moment on the Pr B site is $M_{Pr_B} < 0.1 \cdot M_{Ni_A}$, which is negligibly small. We note that other authors have postulated that the Pr B site has a non-magnetic $4f$ crystal field ground state [19] based on the levels of perovskite $PrNiO_3$ [26]. Additionally, the induced moment for the Pr A site is found to be within ±10% of that of the outer-layer Ni A sites (see *Methods: Density wave models and simulations*). As illustrated in Fig. 5b, for the best fit to the $LowT$ data, both $\boldsymbol{\sigma}_{Ni}$ and $\boldsymbol{\sigma}_{Pr}$ lie parallel rather than perpendicular to $\boldsymbol{q}_{SDW}$ as found in the intermediate phase. We propose below that this is due to a dominant Pr single-ion anisotropy that couples through exchange to the neighboring Ni sublattice (see Discussion and Eq. 1). The corresponding $I^{sim}(\boldsymbol{q})$ is shown in Fig. 4c.

Next, we turn to the $IntW$ phase of $Pr_4Ni_3O_{10}$. Although the overall modulation is similar to the $IntC$ state, significant differences are found in the diffuse peak profiles, indicating that a different set of correlations dominate the scattering upon warming. As illustrated in Fig. 5c, a similar SDW model with a different stacking pattern of the four trilayers in the doubled unit cell along $c$ $[1,0,-1;\ 1,0,-1;\ -1,0,1;\ -1,0,1]$ was found to fit the data better (see Suppl. Section III). Furthermore, the conditions $\boldsymbol{\sigma_{Ni}} \perp \boldsymbol{q_{SDW}}$ and $M_{Ni_B} < 0.6$ found for $IntC$ apply to $IntW$ as well, and a composite model that includes a series of stacking lengths ($L_n < 16$) with a specific $\rho(L_n)$ was needed to fit the profiles. The $\rho(L_n)$ derived from the fitting support a higher density distribution of extended inter-trilayer correlations with $L_n \geq 6$ (see Suppl. Fig. S11), which lends quantitative justification to the claim above that the low-temperature, long-range $c$-axis ordered SDW that results from Ni 3$d$ - Pr 4$f$ exchange pathways has been *imprinted* on the Ni sublattice and retains this imprint even when the Pr sublattice is no longer ordered and the 3$d$-4$f$ coupling is thus quenched.

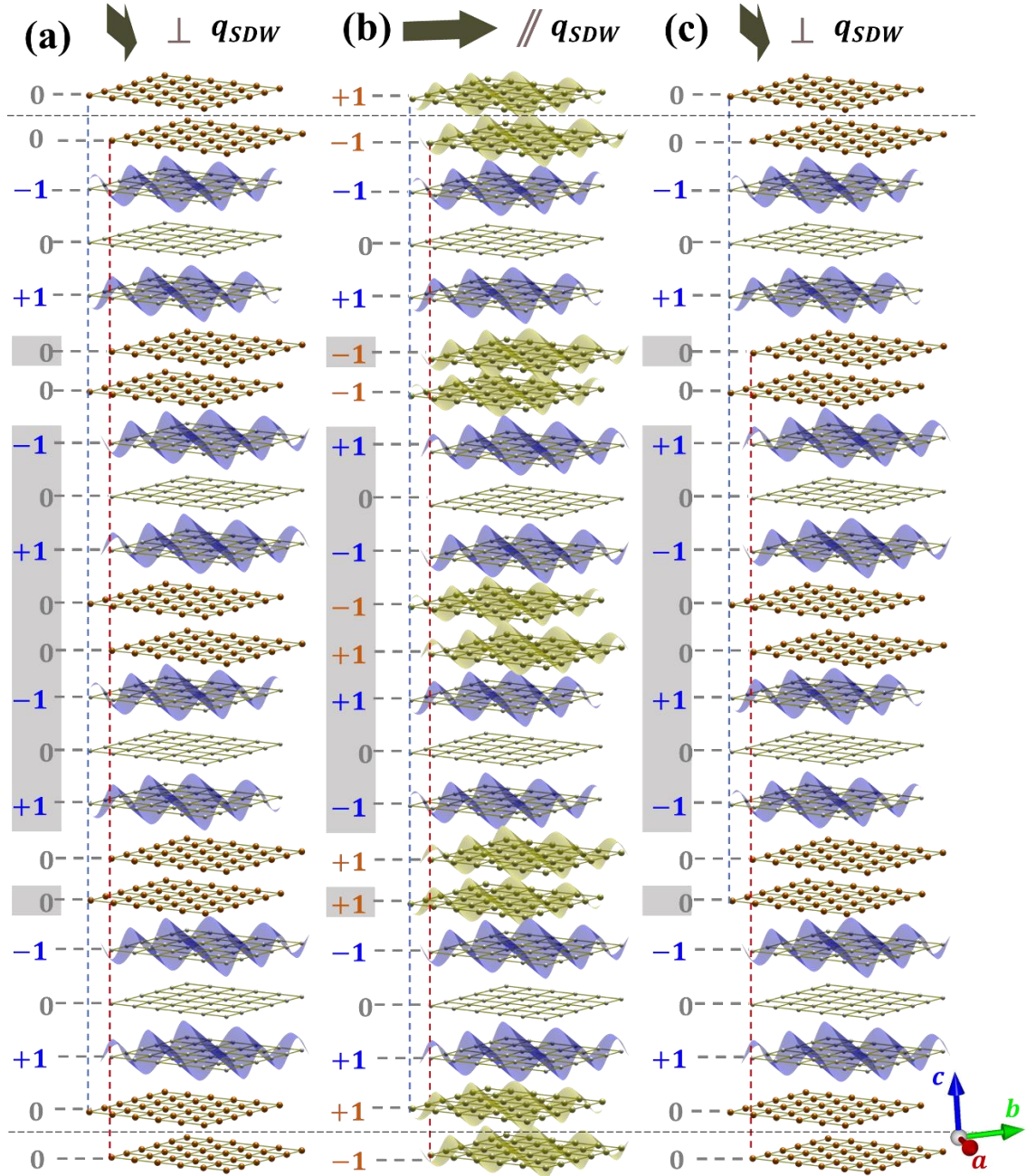

Figure 5: The optimized magnetic models for (a) the intermediate-temperature phase ($T_f < T < T_{MMT}$) while cooling from the paramagnetic phase, $IntC$, (b) the low-temperature phase ($T < T_f$), $LowT$, and (c) the intermediate-temperature phase ($T_f < T < T_{MMT}$) while warming from the low-$T$ phase, $IntW$. The corresponding SDWs are plotted in the orthorhombic $ab$ plane. Note that only Ni A and Pr A layers are shown since moments on Ni B and Pr B are taken as zero. The SDWs shown here are depicted in the same convention as in Ref. [12]. The gray color boxes highlight the in-plane twist operation (twist angle $\varphi$) on the polarization vectors with respect to the rest of the layers (see the Discussion section for more details). Among these three phases, only the $LowT$-phase is long-range ordered, and the other two phases have distributions of stacking lengths as discussed in the text.

To summarize the key results presented above: (1) a quasi-2D ordered SDW in the Ni trilayers with $\boldsymbol{\sigma}_{Ni} \perp \boldsymbol{q}_{SDW}$ induces a corresponding SDW below $T_f \approx 26$ K in the Pr blocks between the trilayers; (2) coincident with this induced order, a 2D to 3D "bootstrapped" crossover leads to long-range 3D order along the $\boldsymbol{c}$-axis for both Ni and Pr sublattices, a doubling of the magnetic unit cell along $\boldsymbol{c}$, and a rotation of the moment such that $\boldsymbol{\sigma} \parallel \boldsymbol{q}_{SDW}$; (3) upon warming, the Pr sublattice loses long-range order at $T_f$, while the Ni sublattice retains a "memory" of the low-$T$ ordering, albeit on a greatly reduced out-of-plane length scale, until the CDW and SDW order simultaneously melt at $T_{MMT} \approx 6T_f$.

## Discussion

Having established the temperature dependence of the magnetic phase behavior and presented models for both the low and intermediate temperature phases consistent with the diffraction data, we now present a phenomenological, microscopic framework that explains all elements of this unusual behavior.

At $T_{MMT}$, Pr$_4$Ni$_3$O$_{10}$ simultaneously develops a CDW and SDW. The CDW correlation length exceeds that of the SDW, which lacks long-range SDW order along $\boldsymbol{c}$ due to the frustrated geometry of the coupling between the trilayers and a weak inter-trilayer exchange expected because of the distance between trilayer blocks. As the temperature is further lowered, the ordering of nickel moments induces a moment on Pr A sites. Due to the lack of mirror symmetry of magnetic Ni sites across the Pr A planes, a net magnetic field will be experienced by the Pr A atoms arising from the Ni SDW order. The local magnetic field on Pr atoms scales with the local moments on nearby nickel sites, and this lifts the Pr A sites out of the crystal field singlet state (see discussion of the crystal field levels in Supplement Section VI) and imparts a SDW onto them with the same basal plane periodicity as the Ni SDW. For this reason, the $k$ component of the magnetic wavevector is the same in both the low and intermediate temperature phases. We conjecture that the Pr B sites lack an induced moment due to the local crystal field inside a perovskite-like trilayer, akin to PrNiO$_3$, for which Pr$^{3+}$ remains non-magnetic [19,26].

For a more quantitative phenomenological understanding, we constructed the following minimal magnetic Hamiltonian:

$$H = \sum_{n=1,2} \sum_{\langle i,j \rangle_n} \boldsymbol{\sigma}_i \cdot J_n \cdot \boldsymbol{\sigma}_j + M_{Pr_A} \sum_{\langle i,j \rangle_3} \boldsymbol{\sigma}_i \cdot J_3 \cdot \boldsymbol{\sigma}_j + M_{Pr_A}^2 \sum_{\langle i,j \rangle_4} \boldsymbol{\sigma}_i \cdot J_4 \cdot \boldsymbol{\sigma}_j \\ - DM_{Pr_A}^2 \sum_{Pr_A} (\boldsymbol{\sigma}_i \cdot \hat{b})^2 \quad (1)$$

$$J_1 = \begin{bmatrix} J_a & 0 & 0 \\ 0 & J_b & 0 \\ 0 & 0 & J_c \end{bmatrix} \quad J_2 = \begin{bmatrix} J_d & 0 & 0 \\ 0 & J_d & 0 \\ 0 & 0 & J_d \end{bmatrix} \quad J_3 = \begin{bmatrix} J_e & 0 & 0 \\ 0 & J_f & 0 \\ 0 & 0 & J_g \end{bmatrix} \quad J_4 = \begin{bmatrix} J_h & 0 & 0 \\ 0 & J_h & 0 \\ 0 & 0 & J_h \end{bmatrix}$$

where $\boldsymbol{\sigma}_i$ is the SDW polarization vector for $i^{th}$ basal plane, and the exchanges $J_n$ are defined as marked in Fig. 1b: $J_1$ and $J_2$ link Ni ions, and $J_3$ and $J_4$ link Ni to Pr and Pr to Pr, respectively, with $M_{Pr}$ defined relative to the Ni A site moments. Note that exchange tensors are defined in a coordinate system where the $x$ axis is parallel to the crystallographic direction $\hat{a}$. Only $J_1$ and $J_3$ are made anisotropic due to the anisotropic nature of the superexchange pathway (see Fig. 1b and Table S3), with $J_1$ enforcing the anti-phasing between the outer Ni layers. Due to the long interaction distance associated with $J_2$ and $J_4$, we treat these as isotropic. In the intermediate phases, $M_{Pr_A} = 0$, and only the $J_1$ and $J_2$ exchanges remain. The exchange parameters should follow simple conditions: $J_a > J_b, J_c \geq 0$ to satisfy $\sigma_{Ni} \perp q_{SDW}$ and $J_d > 0$ to yield the $IntC$ stacking order of the trilayers as the lowest in energy. The experimental observation that $\sigma_{Ni}$ lies along $a$ can be understood from simple considerations of octahedral rotation patterns discussed by Koshibae *et al.* [27]. Due to the anisotropic oxygen environment surrounding Pr A, we introduce the last term to account for possible single ion anisotropy (SIA) with strength $D$ and easy-axis along $\hat{b}$. Further details can be found in Suppl. Section V.

As shown in Figure 5b, the $\sigma$ for the Pr layers (A sites) are experimentally found to be opposite the nearest Ni layer in the same sublattice in the $LowT$ phase. Thus, the $J_3$ exchange interaction should be antiferromagnetic. The $J_4$ exchange connects two consecutive Pr A layers through a nonmagnetic Pr B layer in the same sublattice, and it should be ferromagnetic ($J_h < 0$) with the condition $2|J_d| < |J_h|$ to stabilize the experimentally observed stacking pattern of the $LowT$ phase with a doubled $c$-axis in its magnetic structure. Finally, the easy axis single ion anisotropy (SIA) along $\hat{b} \parallel q_{SDW}$ is introduced to enforce the $LowT$ phase polarization along the SDW propagation direction assuming the $J_3$ tensor follows the same anisotropic condition as $J_1$ ($J_e > \{J_f, J_g\} \geq 0$). Alternatively, the same $LowT$ structure can be stabilized without SIA under the condition $J_f > \{J_e, J_g\} \geq 0$ (see Suppl. Fig. S18).

The difference between the two SDW models for $IntC$ and $IntW$ is that the SDW polarization for half of the basal planes has flipped (compare the gray boxes in Fig. 5a to those of 5b and 5c). The $IntC$ structure should be the global minimum in this temperature regime since it is experimentally found to be the most favorable while cooling from high temperatures. Thus, the $IntW$ structure is a metastable state protected by a sufficiently high energy barrier. An expression can be derived from Eq. 1 for the magnetic energy of the system connecting two solutions in configurational space via a continuous transformation.

$$E \propto -2\left[J_a[\cos^2(\Theta + \varphi) + \cos^2(\Theta)] + J_b[\sin^2(\Theta + \varphi) + \sin^2(\Theta)]\right] + 2J_d[1 - \cos(\varphi)] - 4M_{Pr_A}\left[J_e[\cos^2(\Theta + \varphi) + \cos^2(\Theta)] + J_f[\sin^2(\Theta + \varphi) + \sin^2(\Theta)]\right] - 4J_h M_{Pr_A}^2 \cos(\varphi) - 8D\, M_{Pr_A}^2[\sin^2(\Theta + \varphi) + \sin^2(\Theta)] \quad (2)$$

Here $\Theta$ and $\varphi$ are respectively, the global in-plane rotation of $\sigma$ with respect to the $a$ axis, and the relative rotation of $\sigma$ in adjoining layers (gray boxes in Fig. 5) that for $\varphi = \pi$ takes $IntW$ into $IntC$ (see Suppl. Fig. S16). For illustration, when $\varphi = 0$, $\sigma$ is perpedicular to $q_{SDW}$ for $\Theta = 0$ and $\sigma$ is parallel to $q_{SDW}$ for $\Theta = \pi/2$. Figure 6 shows the energy landscape for a potential choice of exchange parameters as a function of $\varphi$ and $M_{Pr_A}$ at a fixed $\Theta$, where the energy is minimized for each panel. Note that the energy is minimized for $\Theta=0$ for all the presented $M_{Pr_A}$ values except for $M_{Pr_A} = 1$, where it takes the value $\pi/2$. The generalized case of finite $\Theta$ can be found in Suppl. Section V. As shown in the top panel of Figure 6, the global minimum for this Hamiltonian is at $\varphi = 0$ for $\Theta = 0$ ($IntC$). We checked the robustness of this solution under the sensitivity analysis given in Suppl. Fig. S16.

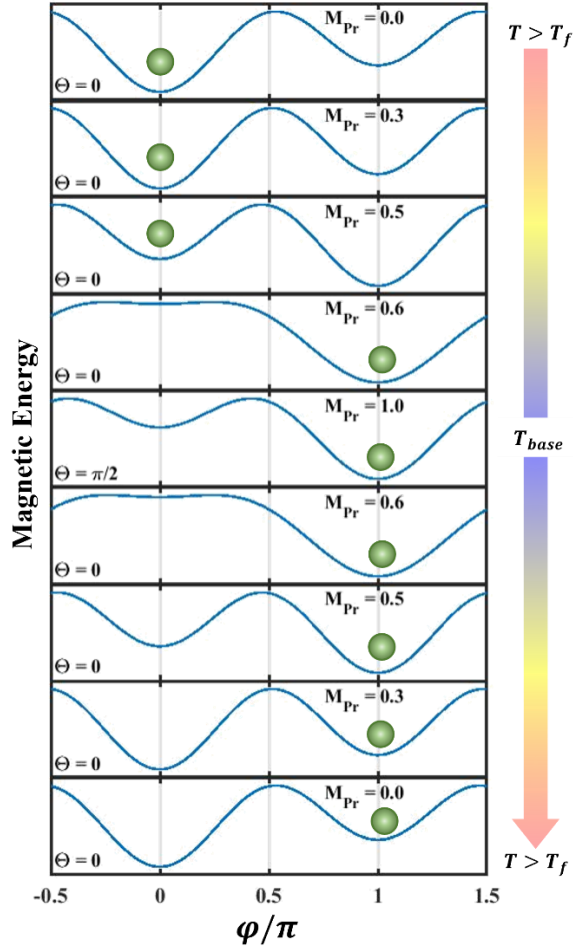

Figure 6: Energy landscape of the two-unit-cell magnetic system in configurational space parameterized by the in-plane twist angle, $\varphi$, with the twisted spins indicated by the gray boxes in Fig. 5. The exchange parameters used here are $J_a = 1, \frac{J_b}{J_a} = 0.5, \frac{J_c}{J_a} = 0, \frac{J_d}{J_a} = 0.1, \frac{J_e}{J_a} = 0.5, \frac{J_f}{J_a} = 0.25, \frac{J_g}{J_a} = 0, \frac{J_h}{J_a} = -0.3$ and $\frac{D}{J_a} = 0.5$. Note that $\varphi = 0$ and $\pi$ correspond to the $IntC$ and $IntW$ structures shown in Fig. 5, respectively. The energy landscapes are derived from Eq. 2 by varying the induced moment on the Pr A site. As shown in Fig. 1e, the induced Pr moment increases with decreasing temperature. In the same notion, $M_{Pr}$ increases from (top) 0 to (middle) 1 and decreases again back to (bottom) 0 to yield the observed thermal hysteresis. The green color markers are a guide to show the stable configuration of the system at a particular step in the process.

The induced moment on the Pr A site is temperature dependent, as observed in the RXS experiment (see Suppl. Fig. S7b). To understand the metastable trapping in the intermediate temperatures upon heating from base temperature, the energy landscape is generated by first increasing $M_{Pr_A}$ (a proxy for temperature) and then decreasing to zero, as shown in Figure 6 from the top panel to the bottom. As $M_{Pr_A}$ increases as a consequence of decreasing temperature, the energy landscape evolves, and after a certain value (in this case $M_{Pr_A} \approx 0.5$), the global minimum is no longer at $\varphi = 0$ but $\varphi = \pi$, which is the *IntW* configuration with low-*T* stacking. Increasing $M_{Pr_A}$ further will drive the system to the *LowT* phase (see Suppl. Fig. S17). On raising *T* (decreasing $M_{Pr_A}$), the system will be moved to the $\varphi = \pi$ configuration and become trapped by the energy barrier between the $\varphi = 0$ and $\pi$ configurations. Apparently, this barrier is sufficiently high in practice to protect the metastable *IntW* state until all order vanishes at $T_{MMT}$. Our minimal model thus captures all experimentally observed elements, including the rotation of *σ*, the metastability of the *IntW* phase, and the thermal hysteresis of the Ni-centered SDW.

Let us consider the case of $Pr_4Ni_3O_{10}$ within the context of other transition metal oxides where both transition metal and rare-earth sublattices are present. Considerable experimental work has been done on the rare-earth magnetism in layered cuprates. In the 1-2-3, 2-4-8, and 2-4-7 cuprates, commensurate antiferromagnetic rare-earth ordering is observed in the materials containing doubly-degenerate rare earth *4f* crystal field states, the ordering occurring at low temperatures ($\approx$ K) due to dipole-dipole interactions [28]. This is exemplified by $ErBa_2Cu_3O_7$, the Er sublattice behaving as a realization of the 2D *S*=1/2 Ising model [28]. Well below $T_N$, i.e. << 1 K, the rare earth sublattice can evolve from 2D to 3D due to even a weak *c*-axis coupling that gets magnified by the sheer number of interacting spins within a domain to lower the energy [28], as occurs in $Er_2Ba_4Cu_8O_{16}$ [29]. The extremely low temperatures at which this latter effect occurs reflects a different mechanism than described here. The single-layer T' 2-1-4 cuprate, $Nd_2CuO_4$, provides an interesting comparison to $Pr_4Ni_3O_{10}$. The Cu spins develop antiferromagnetic order at 245 K; a spin reorientation occurs at $\approx$ 30 K, followed by an induced moment on the Nd sublattice [30]. The reorientation and the smeared nature of the low temperature order parameter are reminiscent of the induced moment in $Pr_4Ni_3O_{10}$. However, the dimensional crossover and imprinting effects have not been reported in any cuprates that we are aware of. In nickelates, neutron diffraction shows that the Ni sublattice $Pr_2NiO_4$ orders antiferromagnetically at 325 K (where the crystal symmetry is *Bmab*) with moments along the *a* axis, as we find in $Pr_4Ni_3O_{10}$ [27,31]. The low temperature magnetic behavior of $Pr_2NiO_4$ is more complex than that of $Pr_4Ni_3O_{10}$, including spin reorientations and the emergence of a ferromagnetic component allowed by a structural phase transition at 115 K to $P4_2/ncm$. Below 40 K, an increase in magnetization has been associated to partial ordering of $Pr^{3+}$, with neutron diffraction intensities attributed to polarization of $Pr^{3+}$ by the exchange field of the ordered Ni sublattice. However, neither a 2D-3D crossover nor feedback to the Ni sublattice has been reported in $Pr_2NiO_4$. It is worth noting that non-Kramers $Pr^{3+}$ in both $Pr_2NiO_4$ and in $Pr_4Ni_3O_{10}$ should be in a singlet ground state due to their low site-symmetry. Thus, a magnetic transition on the Pr sublattice as proposed by Rout *et al.* [19] seems unlikely (see Suppl. Section VI).

## Summary

In summary, we have used a combination of neutron diffraction and resonant X-ray diffraction to unveil the evolution of the SDW, which is coupled to a CDW, in $Pr_4Ni_3O_{10}$. Detailed models of the magnetic ordering in all phases have been presented, and a microscopic model of exchange interactions is proposed to explain the phase behavior and the metastability of the intermediate temperature state on warming. The first striking experimental finding from this work is that the dimensionality of the magnetic correlations in the SDW is modified as a function of temperature by the interaction of the Pr 4$f$ moments with the Ni 3$d$ moments. Above 26 K, the Ni sublattice orders in 2D, with weak correlations along $c$ enforced by geometric frustration. Below 26 K, a new Pr-centered order develops, whose temperature dependence suggests an induced-moment effect from the Ni sublattice on what would otherwise be non-magnetic, singlet state ions. Once a finite ordered moment is induced on $Pr^{3+}$, a new exchange pathway opens along $c$, breaking the frustration and leading to 3D order of both sublattices. Given that these cascading events originate from the order of the Ni cations themselves, we describe this mechanism as bootstrapping (a phenomenon known colloquially in physics as "pulling oneself up by the bootstraps"). Remarkably, the imprint of this effect on the nickel sublattice survives a re-heating back into the intermediate-temperature phase and only disappears if the temperature is increased above $T_{MMT}$, where the CDW/SDW melts. Surprisingly, we see no evidence that the CDW is modified by the dramatic evolution in the SDW, despite the obvious coupling between the SDW and CDW (Fig. 1c). Specifically, we observed no significant difference in the in-plane wave-vector or the line-shape of the CDW between the intermediate and low temperature states (see Suppl. Fig. S6). It is possible that such impact in the charge sector is too subtle for the diffraction probes used here, and we might expect to find a spectroscopic signature of the coupling using other probes.

It would be surprising if $Pr_4Ni_3O_{10}$ were a singular example of the phenomena described here. Nor would the bootstrapping effect need to be confined to the magnetic sector due to the strong coupling among structural, electronic and magnetic degrees of freedom commonly found in low-dimensional quantum materials. Along this line, we note that the substitution of Pr cations into YBCO has recently been claimed to precipitate a 3D charge ordered state due to hybridization of the Pr 4$f$ states with the oxygen 2$p$ states [32]. Although this occurs in the charge sector and doesn't yield a temperature-dependent cross-over as seen here, it points to another example where rare earth cations can be used to control the dimensionality of density waves. We hope that the interesting case presented here will inspire work on further investigations of 4$f$-3$d$ electron interactions in layered oxides to probe and understand the interactions that underlie other novel dimensional crossover phenomena.

## Methods

**Sample growth and characterization.** Description of the growth of the Pr$_4$Ni$_3$O$_{10}$ and La$_4$Ni$_3$O$_{10}$ crystals employed in this study, as well as the electrical transport characterization of Pr$_4$Ni$_3$O$_{10}$, can be found in our previous report [17].

**Resonant X-ray scattering.** Synchrotron X-ray single-crystal diffraction measurements were performed at Sector 6-ID-B at the Advanced Photon Source, Argonne National Laboratory. The resonant measurements were made with incident X-ray energies around the Pr $L_2$-edge at 6.439 keV and the Ni K-edge at 8.333 keV. For the former measurements, a polarization analyzer in $\sigma - \pi'$ configuration with the vertical scattering geometry was used. The magnetic reflection $\boldsymbol{Q}_{RXS} = $ (0, -0.617, 9.5) was measured along [00$l$] as a function of the azimuthal angle ($\Psi$) at a fixed temperature of 5 K.

**Neutron scattering.** Elastic neutron scattering experiments were carried out on a Pr$_4$Ni$_3$O$_{10}$ crystal at the MACS triple-axis spectrometer and the SPINS triple-axis spectrometer [33], both located at the NIST Center for Neutron Research, Gaithersburg, MD. For both spectrometers, the incident and final neutron energies were fixed to be $E_i = E_f = $ 5.0 meV, and the harmonics were suppressed by beryllium filters. Further neutron investigations were performed on the same Pr$_4$Ni$_3$O$_{10}$ crystal as well as a single crystal of La$_4$Ni$_3$O$_{10}$ at the diffuse scattering spectrometer CORELLI, Spallation Neutron Source, Oak Ridge National Laboratory. CORELLI is a time-of-fight instrument where the elastic contribution is separated by a pseudo-statistical chopper [34]. The crystal was rotated through 360 degrees with the step of 3° horizontally with the vertical angular coverage of ±28.5° for survey of the elastic and diffuse peaks in reciprocal space. A standard closed cycle refrigerator (CCR) accessible at the CORELLI beamline was used. The data were reduced using Mantid [35] and Python scripts available at CORELLI.

**Density wave models and simulations.** The neutron scattering cross-section for a given SDW model is calculated using:

$$I(\boldsymbol{Q}) = \sum_{\alpha,\beta} \frac{g_\alpha g_\beta}{4} \left( \delta_{\alpha\beta} - \frac{\kappa_\alpha \kappa_\beta}{|\boldsymbol{Q}|^2} \right) \times S^{\alpha\beta}(\boldsymbol{Q}) \qquad (3)$$

where $g_{\alpha,\beta}$ are the $g$-factors, $\boldsymbol{Q} = \kappa_x \hat{x} + \kappa_y \hat{y} + \kappa_z \hat{z}$ is the wavevector transfer in the scattering process, $\alpha, \beta = x, y, z$ are the initial and final spin polarizations of the neutrons in cartesian coordinates, and $S^{\alpha\beta}$ is the neutron scattering factor correlation function:

$$S^{\alpha\beta}(\boldsymbol{Q}) = \frac{1}{2\pi N} \left| S_{\boldsymbol{Q}}^\alpha S_{-\boldsymbol{Q}}^\beta \right| \qquad (4)$$

with

$$S_{\boldsymbol{Q}}^\alpha = \sum_i F_i(\boldsymbol{Q}) \cdot \sigma_i^\alpha \cdot \cos(\boldsymbol{q}_{SDW} \cdot r_i + \phi) \, e^{i\boldsymbol{Q} \cdot r_i} \qquad (5)$$

Unlike in systems with localized magnetic moments, which is usually represented as $S_i$ with fixed amplitude, here the effective moment per ion is decomposed into a cosine component $\cos(\boldsymbol{q_{SDW}} \cdot r_i + \phi)$ characterized by the SDW propagation vector, $\boldsymbol{q_{SDW}}$ and the SDW polarization $\boldsymbol{\sigma}_i^\alpha$ to capture both the itinerant (single harmonic) and localized (Ni-site centered) behavior. The offset parameter $\phi$ is considered constant (0 or $\pi$) for all the basal planes in each orthorhombic sublattice shown in Suppl. Fig. S14. $F_i(\boldsymbol{Q})$ is the magnetic form factor of the $i^{th}$ magnetic site. To accurately account for the instrumentation effects, $I(\boldsymbol{Q})$ was convoluted with a Lorentzian of FWHM = 0.027 Å$^{-1}$ estimated from the strong nuclear Bragg peaks.

Diffuse scattering contributions due to short-range ordering of Ni-trilayers was calculated using $L_n/2$ unit cells along $\boldsymbol{c}$ ($L_n$ - number of trilayers) and 2500 orthorhombic unit cells in the plane, which is equivalent to a total of 15000× $L_n$ Ni ions. The slight *Bmab* deviations of the outer plane Ni A ions from their *I4/mmm* positions were taken into account, using atomic coordinates given in Ref. [17]. $I(\boldsymbol{Q})$ for simple combinations of $L_n$ stackings are shown in Suppl. Fig. S9.

For the purpose of optimization, each Ni trilayer was parameterized by three local rotation angles $\alpha^n_{\text{Ni}_A}, \beta^n_{\text{Ni}_A}, \gamma^n_{\text{Ni}_B}$ ($n = 1, 2, ...$) and one global moment amplitude ($M_{\text{Ni}} = M_{\text{Ni}_B}/M_{\text{Ni}_A}$) as described in Suppl. Fig. S8. In addition, the probability density for each stacking of length $L_n$ was also refined for the models that allow multiple stackings to exist simultaneously. For each Ni-trilayer, the conditions that the outer layer polarizations, $\boldsymbol{\sigma}_{\text{Ni}_A}$ are equal and opposite, and the middle layer polarization, $\boldsymbol{\sigma}_{\text{Ni}_B}$, is perpendicular to $\sigma_{\text{Ni}_A}$, are enforced.

From the initial optimization attempts using $IntC$ data and simple SDW models that accommodate up to two types of stacking (see Suppl. Section III), We realized that the optimal solution lies at $\alpha^n_{\text{Ni}_A} \approx \pi/2$, $\beta^n_{\text{Ni}_A} \approx 0$ or $\pi$, $M_{\text{Ni}} < 0.6$ and $\gamma^n_{\text{Ni}_B} \in [0, 2\pi]$ (unrefinable for lower $M_{\text{Ni}}$), and $\{\alpha^n_{\text{Ni}_A}, \beta^n_{\text{Ni}_A}, \gamma^n_{\text{Ni}_B}, M_{\text{Ni}}\}$ is fixed to $\{\pi/2, 0, 0, 0\}$ for longer range SDW models with $L_n$ up to 16 (here, $M_{\text{Ni}}$ refers to the Ni B sites). For the SDW models corresponding to the Pr$_4$Ni$_3$O$_{10}$ $IntC$ structure and La$_4$Ni$_3$O$_{10}$ low-$T$ structure, the stackings followed the pattern {[1, 0, -1], [1, 0, -1], [1, 0, -1], [1, 0, -1],…}. ([…] represents a single trilayer). In contrast, the Pr$_4$Ni$_3$O$_{10}$ $IntW$ model followed the pattern {[1, 0, -1], [1, 0, -1], [-1, 0, 1], [-1, 0, 1],…}. Note that stackings with $L_n < 3$ are the same for both models. The refined probabilities as a function of $L_n$ are given in Suppl. Fig. S11.

For the purpose of calculating $I(\boldsymbol{Q})$ for a long-range stacking of the SDW, a box of 20×20×10 nuclear unit cells, which is equivalent to a total of 224,000 magnetic ions including both Ni and Pr sites, was considered along with periodic boundary conditions.

In the low-$T$ data-fitting process, the conditions of outer Ni-layer polarizations within a trilayer block being equal and opposite, and the corresponding middle layer polarization being perpendicular to the outer layers, is enforced. The magnetic unit cell was assumed to be doubled along $\boldsymbol{c}$ (see Fig. 5), with the parameterization referring to the bottom unit-cell (see Suppl. Fig. S8). The induced SDWs for Pr are considered similar to the Ni-layers (same $\boldsymbol{q_{SDW}}$ and $\phi$, not $\boldsymbol{\sigma}$), which belong to the same sublattice as shown in Suppl. Fig. S14. However, each Pr layer is parameterized by two local rotation angles $\alpha^n_{\text{Pr}_A}, \beta^n_{\text{Pr}_A}$ ($n = 1 ... 8$) and two global amplitudes:

$M_{Pr_A}$, $M_{Pr_B}$ — the magnetic moment of Pr A and B sites relative to the outer layer Ni$_A$ moment, respectively. The optimal region for the Pr$_4$Ni$_3$O$_{10}$ *LowT* neutron and RXS data is found to be at $M_{Pr_A} \approx 1$, $M_{Pr_B} \approx 0$, $\alpha_{Pr_A} = \pi$, $\beta_{Pr_A} = 0$, $\alpha^2_{Ni_A} = \beta^1_{Ni_A} = \beta^2_{Ni_A} = 0$ and $\alpha^1_{Ni_A} = \pi$. Similarly, in diffuse scattering data fitting, $M_{Ni} < 0.6$ is found (i.e., Ni B sites). Since the Pr B site and middle layer Ni$_B$ moments are small, the corresponding parameters for their orientation ($\gamma^1_{Ni_B}, \gamma^2_{Ni_B}, \alpha_{Pr_B}$ and $\beta_{Pr_B}$) are unconstrained and set to zero in the final solution shown in Fig. 4c.

**Optimization.** To quantify the uncertainty of the proposed SDW models, we applied the iterative optimization procedure (IOP) explained in Ref. [36]. In the case of diffuse scattering data fitting, the direct distance between the experimentally measured ($I^{exp}(\boldsymbol{Q})$) and the calculated ($I^{sim}(\boldsymbol{Q})$) neutron structure factors, $\chi^2 = \frac{1}{N_Q}\sum[I^{exp}(\boldsymbol{Q}) - I^{sim}(\boldsymbol{Q})]^2$ was employed as the cost function of IOP.

In the case of *LowT* data fitting, a composite cost combining neutron data and RXS data,

$$\chi^2 = \frac{1}{N_{Q_P}}\sum_{\boldsymbol{Q}_P}[I^{exp}(\boldsymbol{Q}_P) - I^{sim}(\boldsymbol{Q}_P)]^2 + \frac{1}{N_\Psi}\sum_{\Psi}[I^{exp}_{RXS}(\boldsymbol{Q}_{RXS}, \Psi) - I^{sim}_{RXS}(\boldsymbol{Q}_{RXS}, \Psi)]^2 \quad (6)$$

where $I_{RXS}$ is the Pr L$_2$ edge X-ray resonant scattering cross-section at the wavevector $\boldsymbol{Q}_{RXS} = [0, -0.6, 9.5]$. (see Suppl. Section II for details) and $\boldsymbol{Q}_P$ represents the SDW neutron peaks.

## Acknowledgements

This work was supported by the US Department of Energy, Office of Science, Basic Energy Sciences, Materials Science and Engineering Division. Research at ORNL's SNS was sponsored by the Scientific User Facilities Division, Office of Basic Energy Sciences, U.S. Department of Energy (DOE). Use of the Advanced Photon Source at Argonne National Laboratory was supported by the U.S. Department of Energy, Office of Science, Office of Basic Energy Sciences, under Contract No. DE-AC02-06CH11357. Access to MACS was provided by the Center for High Resolution Neutron Scattering, a partnership between the National Institute of Standards and Technology and the National Science Foundation under Agreement No. DMR-1508249. Junjie Zhang gratefully acknowledges financial support from the National Natural Science Foundation of China (Grant No. 12074219), the Qilu Young Scholars Program of Shandong University, and the Taishan Scholars Program of Shandong Province.

# Supplementary Information:

# Bootstrapped Dimensional Crossover of a Spin Density Wave

Anjana Samarakoon[1], J. Strempfer[2], Junjie Zhang[1,3], Feng Ye[4], Yiming Qiu[5], J.-W. Kim[2], H. Zheng[1], S. Rosenkranz[1], M. R. Norman[1], J. F. Mitchell[1] and D. Phelan[1]

[1]*Materials Science Division, Argonne National Laboratory, Lemont, IL 60439, USA*

[2]*Advanced Photon Source, Argonne National Laboratory, Argonne, Illinois 60439, USA*

[3]*State Key Laboratory of Crystal Materials, Shandong University, 250100, Jinan, Shandong, China*

[4]*Neutron Scattering Division, Oak Ridge National Laboratory, Oak Ridge, TN 37830, USA*

[5]*NIST Center for Neutron Research, National Institute of Standards and Technology, Gaithersburg, Maryland 20899, USA*

This PDF file includes:

I.      Additional scattering experiments and data analysis

II.     Resonant X-Ray Scattering experiment

III.    Modeling and optimization of diffuse scattering phase

IV.    Modeling and optimization of low temperature phase

V.     Phenomenological model and analysis

VI.    Pr Crystal Electric Fields and Induced Magnetism

Figures S1-S19

# I. Additional scattering experiments and data analysis

Figure S1 shows the untreated (raw) data corresponding to the linecuts shown in Fig. 3 and Fig. 4 from neutron diffuse scattering experiments performed at the CORELLI spectrometer, SNS, Oak Ridge National Laboratory. These linecuts are made out of volumetric data along $l$ by integrating over a range of $h \pm 0.1$ and $k \pm 0.1$.

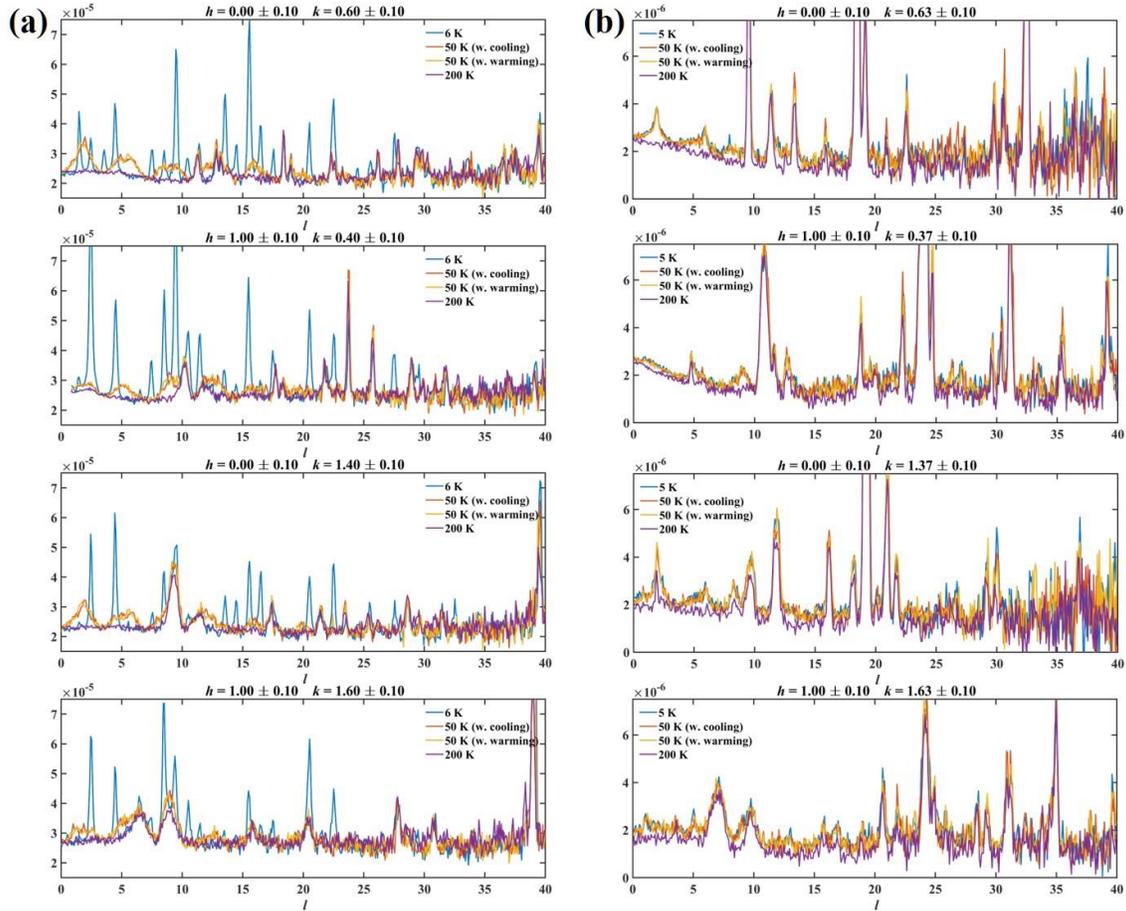

Figure S1: Raw data from the CORELLI experiment for (a) $Pr_4Ni_3O_{10}$ and (b) $La_4Ni_3O_{10}$ are shown here. Each panel contains four sub-panels for four different combinations of $h \pm 0.1$ and $k \pm 0.1$ and each sub-panel contains line cuts along $l$ for the measurement taken at base-temperature, $T_{base}$ (5 K or 6 K), 50 K after cooling from 200 K, 50 K after warming from $T_{base}$ and 200 K. The 200 K data were treated as the non-magnetic background and subtracted before the modeling as shown in Fig. 4.

2D Slices of the background-subtracted CORELLI data collected on Pr$_4$Ni$_3$O$_{10}$ at 6 K and 50 K (after cooling from 200 K) are shown in Fig. S2. The slices are made in the $(hk0)$ plane by integrating over a range of $l \pm 0.25$. Note that the magnetic peaks are significantly narrower along $h$ and $k$ than along $l$, and they lack the obvious diffuse scattering that was observed in the $(0kl)$ plane as described in the text. Even though the SDW peaks show $C_4$-symmetry, we attribute this to the presence of two orthogonal nuclear domains (twins). The two sets of magnetic peaks are marked using open circles and squares in Fig. S2 (a). Prior X-ray diffraction studies on La$_4$Ni$_3$O$_{10}$ concluded that the superlattice (SL) peaks from a single-domain crystal are either along $h$ or $k$, depending on the crystal orientation. Moreover, because *a* and *b* differ by <1%, the crystals are susceptible to twinning.

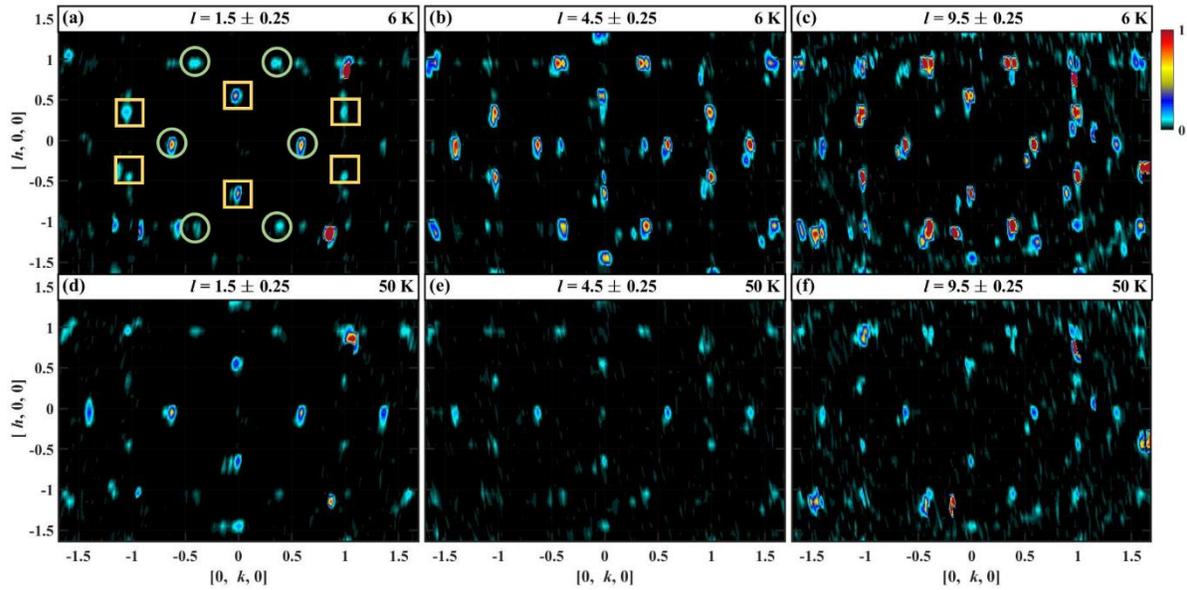

Figure S2: The $(hk0)$ slices of CORELLI data collected on Pr$_4$Ni$_3$O$_{10}$ for (a, d) $l = 1.5 \pm 0.25$, (b, e) $l = 4.5 \pm 0.25$ and (c, f) $l = 9.5 \pm 0.25$. The data shown here were collected at (a-c) $T = 6$ K and (d-f) $T = 50$ K. The two sets of incommensurate peak structures corresponding to two orthogonal nuclear domains (twins) are identified and marked as either open-circles or open-squares in panel (a). (see Fig. S3) Only the domain shown by the open-square set was accounted for in the SDW modeling.

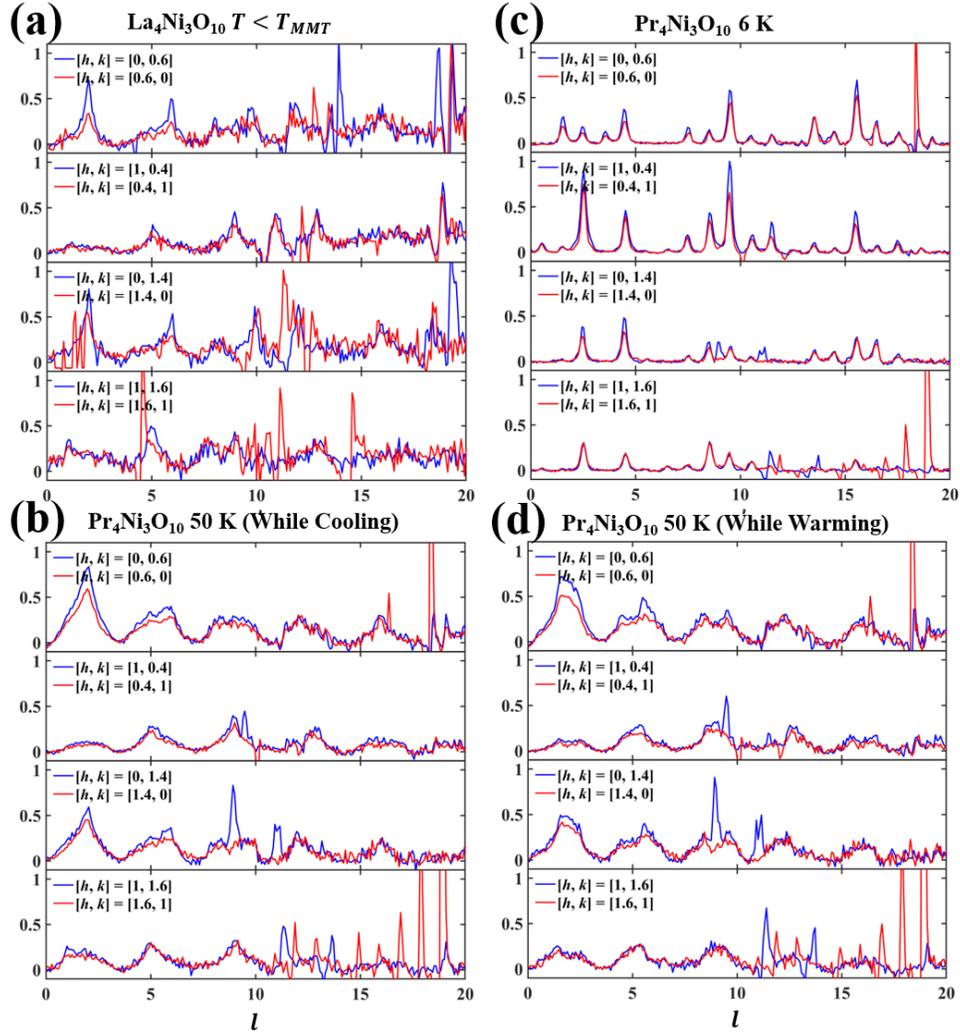

Figure S3: Comparison of neutron data collected on equivalent peaks of $[h, k]$ from equivalent orthogonal nuclear domains (twins). (a) $La_4Ni_3O_{10}$ data collected at $T < T_{MMT}$, (b) $Pr_4Ni_3O_{10}$ data at 50 K measured while cooling from 200 K, (c) $Pr_4Ni_3O_{10}$ data collected at base-temperature (~6 K) and (d) $Pr_4Ni_3O_{10}$ data measured at 50 K while warming from base-temperature. Because of the temperature independent nature of the structure factors below $T_{MMT}$ for $La_4Ni_3O_{10}$ (see Fig. 3d), all the low temperature data were combined in panel (a). Each panel shows line cuts of 3D scattering data along $l$ for four different combinations of $h \pm 0.1$ and $k \pm 0.1$: [0, 0.6], [1, 0.4], [0, 1.4], and [1, 1.6] in comparison to corresponding equivalent peaks of [0.6, 0], [0.4, 1], [1.4, 0], and [1.6, 1].

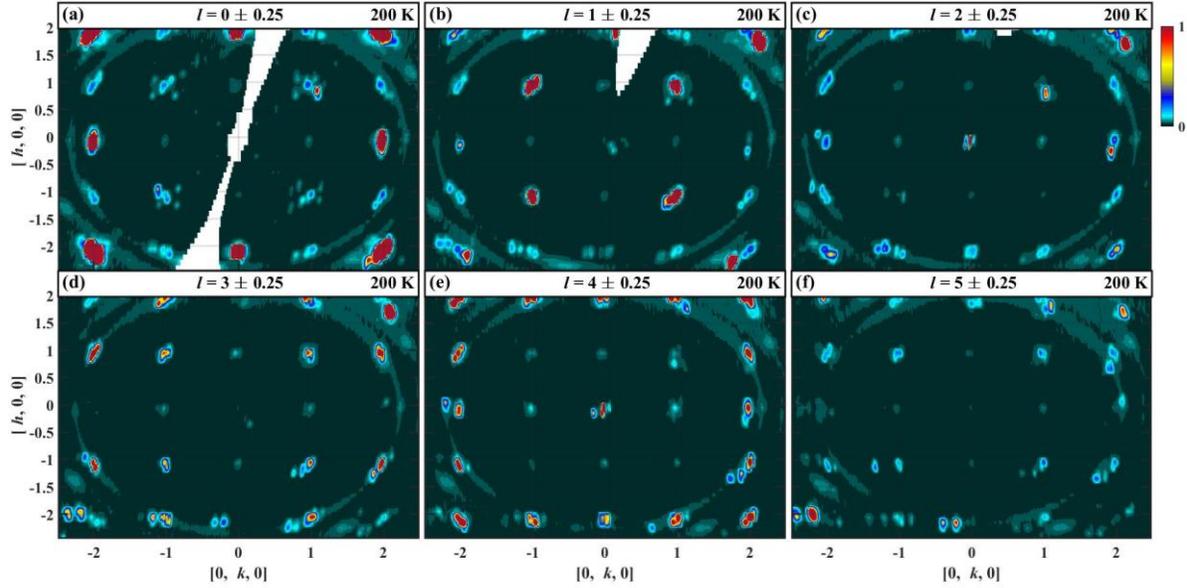

Figure S4: The $(hk0)$ slices of CORELLI data collected on $Pr_4Ni_3O_{10}$ at $T = 200$ K for (a) $l = 0 \pm 0.25$, (b) $l = 1 \pm 0.25$, (c) $l = 2 \pm 0.25$, (d) $l = 3 \pm 0.25$, (e) $l = 4 \pm 0.25$ and (f) $l = 5 \pm 0.25$.

**Temperature dependence of spin density wave (SDW) peaks**

The evolution of the temperature dependent magnetic correlations along $c$ is most easily observed by scans at constant $h$ and $k$ and varying $l$. Figure S5 shows such measurements taken on the SPINS triple-axis spectrometer along $[0, 0.605, l]$ over a range of $l$ varying from 0 to 3 at multiple temperatures between 4.5 K and 180 K. The scan at 180 K shows a relatively constant background, whereas magnetic scattering is observed at the scans taken below 160 K.

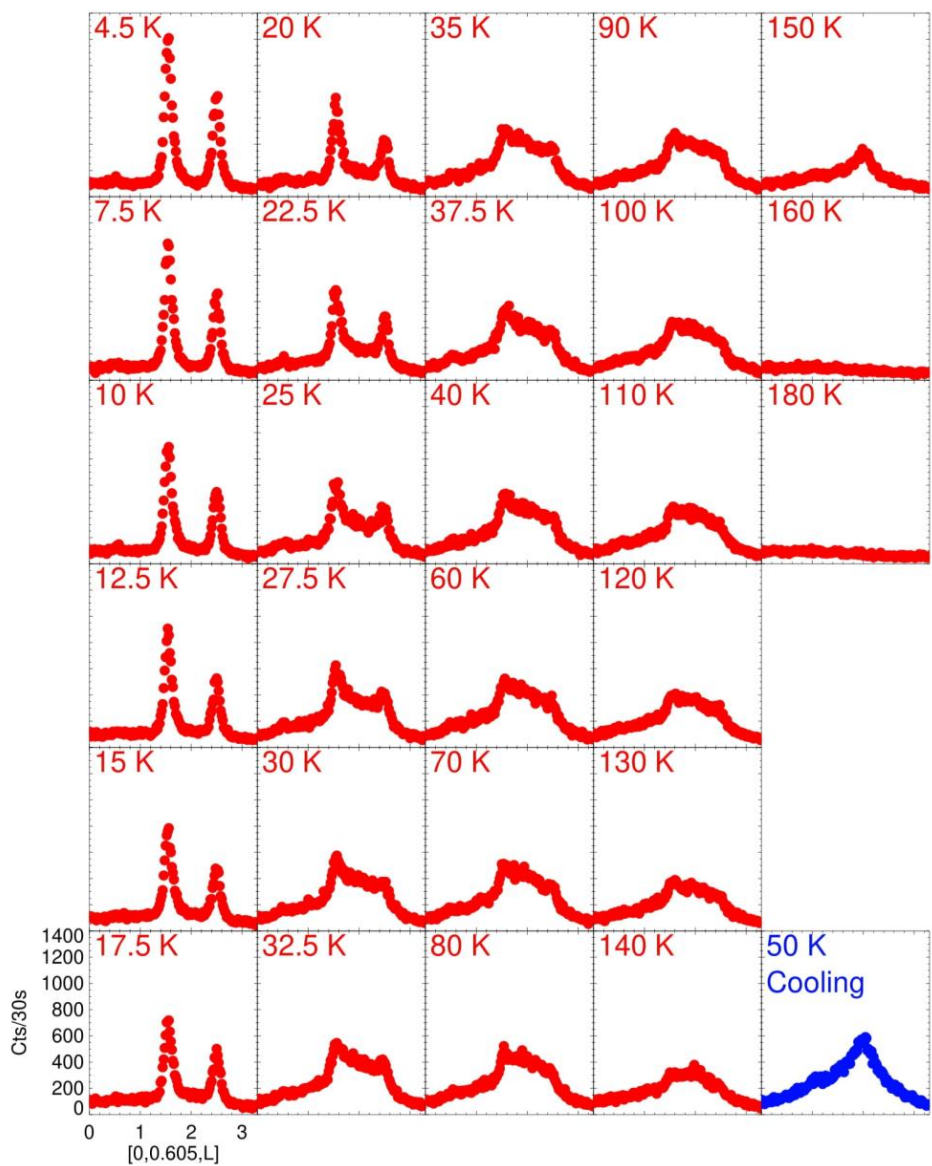

Figure S5: The data collected on $Pr_4Ni_3O_{10}$ along $[0, 0.605, l]$ as a function of temperature using the SPINS triple-axis spectrometer. Throughout the experiment, the sample was oriented in the $(0kl)$ scattering plane. Except for 50 K data set, the sample was first cooled down to a base temperature of 4.5 K, and data were collected while warming. The 50 K data (blue) were collected after cooling from a higher temperature. The change of peak profile of 50 K versus 40 K or 60 K is due to the thermal hysteresis effect, which agrees with our observations from the CORELLI experiments.

**Temperature dependence of the charge density wave (CDW)**

Figure S6 shows the CDW peak at (1,-2.23,8) measured for $Pr_4Ni_3O_{10}$ as a function of temperature. Unlike its magnetic counterpart, the CDW order is robust below $T_{MMT} \approx 158\ K$. The resulting order parameter extracted from this peak (integrated area) as a function of $T$ is shown in Fig. 1e.

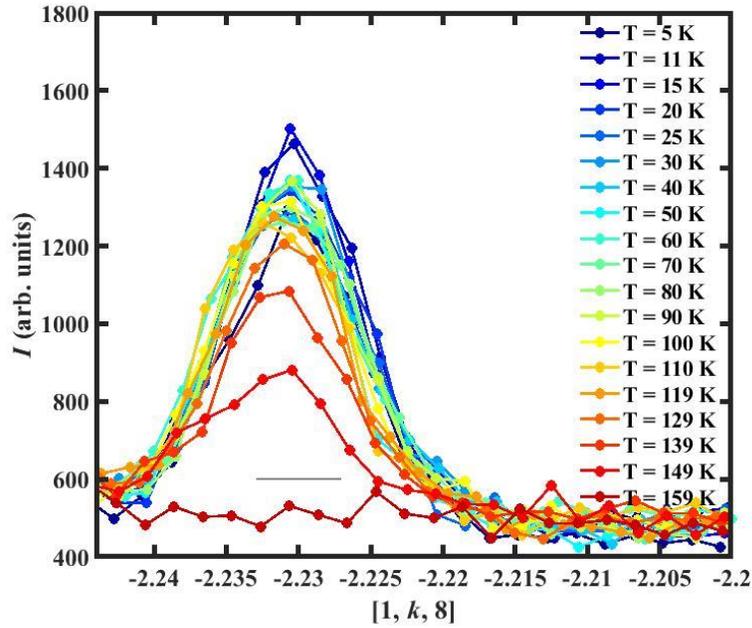

Figure S6: The temperature evolution of the CDW peak (1, -2.23, 8) measured using single-crystal synchrotron x-ray diffraction. The instrument resolution (FWHM of a nuclear Bragg peak) is also shown as a gray horizontal line.

## II. Resonant X-Ray Scattering experiment

Resonant X-ray scattering (Pr $L_2$ edge) was measured on the peak at $\boldsymbol{Q}_{RXS} = (0, -0.617, 9.5)$ at different values of the azimuthal angle ($\Psi$) as shown in Suppl. Fig. S7a. The peak intensity as a function of $\Psi$, extracted in a post-processing step, is shown in Fig. 2d.

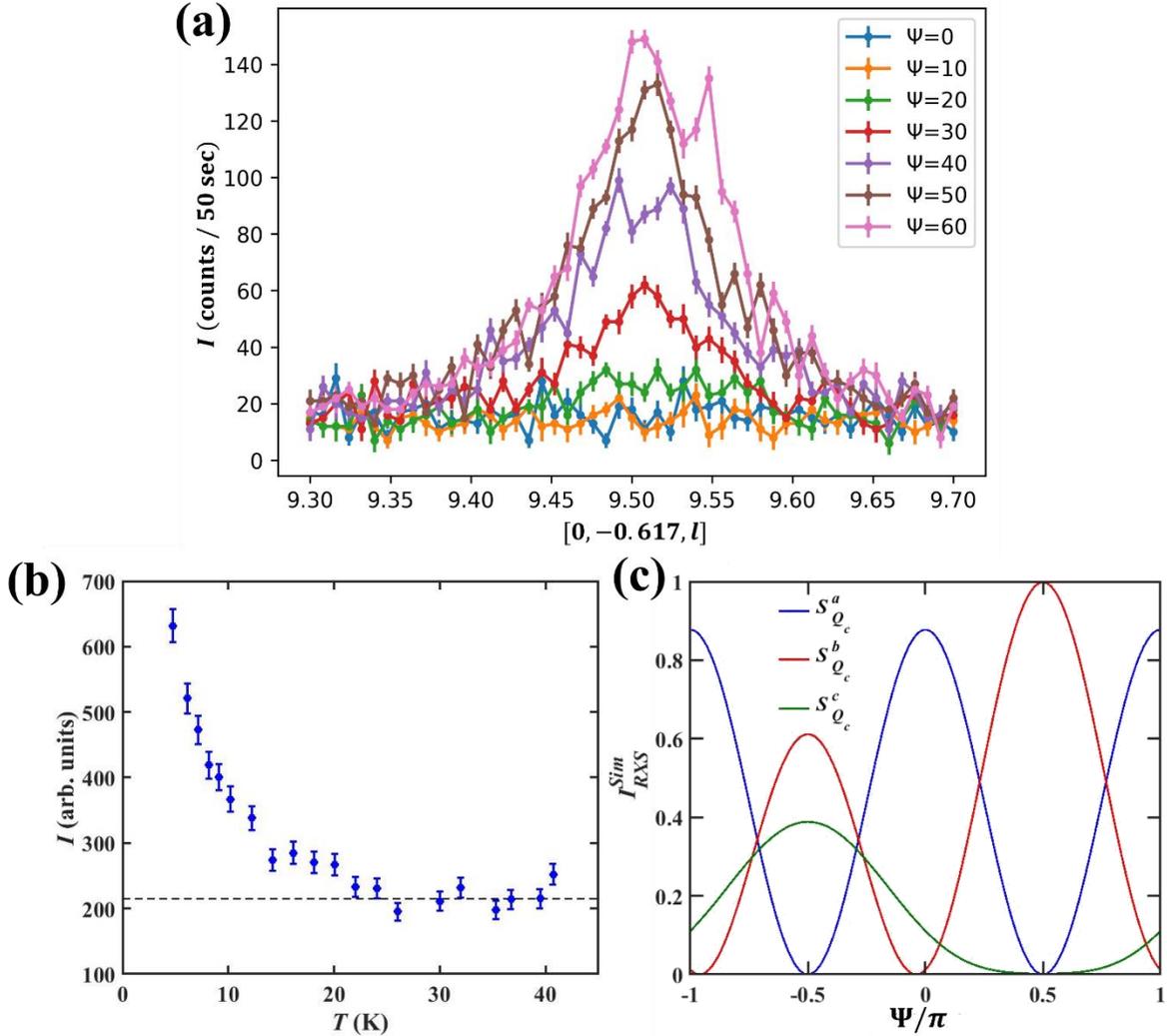

Figure S7: Resonant X-Ray scattering experiments, analysis, and modeling at the Pr $L_2$ edge. (a) Azimuthal ($\Psi$) dependence of the magnetic peak at $\boldsymbol{Q}_{RXS} = (0, -0.617, 9.5)$. (b) Temperature dependence of the ordered Pr (magnetic) moment at a fixed azimuthal angle ($\Psi = -53$ deg). (c) The calculated $\Psi$ dependences of the structure factor $I_{RXS}^{Sim}$ for the Pr-only spin-correlations at $\boldsymbol{Q}_{RXS}$ being along the crystallographic axes $\boldsymbol{a}$, $\boldsymbol{b}$ and $\boldsymbol{c}$. The red curve is most consistent with the data.

**Calculating the azimuthal dependance**

The azimuthal dependence of the Pr $L_2$ edge X-ray resonant scattering cross-section is calculated using

$$I_{RXS}^{Sim}(\Psi) = [\cos(A)\sin(\theta) - \sin(A)\sin(B + \Psi)\cos(\theta)]^2 \tag{S1}$$

$$A = \cos^{-1}(S_{Q_c} \cdot U_3)$$

$$B = \tan^{-1}\left(\frac{S_{Q_c} \cdot U_1}{S_{Q_c} \cdot U_2}\right)$$

where $S_{Q_{RXS}} = [S_{Q_{RXS}}^x, S_{Q_{RXS}}^y, S_{Q_{RXS}}^z]$ is the Pr contribution to the spin density at the wavevector $Q_{RXS} = (0, -0.617, 9.5)$ (the Bragg angle for the experiment is $\theta = 0.114\pi$) and $U_1$, $U_2$, $U_3$ have the same definitions as in Ref. [1], with Equation S1 derived from Eq. 15 of Ref. [1]. $I_{RXS}^{Sim}(\Psi)$ for principal directions of $S_{Q_{RXS}}$ is shown in Fig. S7c.

### Charge density wave (CDW) peaks at the Ni *K*-edge

Scans of the X-ray energy were performed on $Q = (0, -1.23, 17)$ and off $Q = (0, -1.11, 17.2)$ a CDW peak through the Ni *K*-edge as shown in Fig. 2a. The data here have been scaled to the maximum intensity to facilitate the comparison of the line shapes. The shift to lower energy of the CDW peak between the two $Q$ indicates that the Ni cations contribute to the structure factor of the CDW peak.

## III. Modeling of the diffuse scattering data

Figure S8a illustrates the SDW model and the corresponding parameterization used to fit the diffuse neutron scattering data from the $Pr_4Ni_3O_{10}$ intermediate phase and the $La_4Ni_3O_{10}$ *LowT* phase. Note that the wavy surfaces on each plane have a different meaning from Fig. 5, in that it is merely the cosine modulation $\cos(\boldsymbol{q_{SDW}} \cdot r_i + \phi)$ from Eq. 5 instead of $\boldsymbol{\sigma}_i^\alpha \cdot \cos(\boldsymbol{q_{SDW}} \cdot r_i + \phi)$ as depicted in Fig. 5 of main text or Fig. 4 of Ref. [2]. The black arrows shown on the right side of each plane are the local SDW polarization directions $\boldsymbol{\sigma}_i$. The structure factor, $I(\boldsymbol{Q})$, calculation and the SDW model are explained in *Methods: Density wave models and simulations*. An iterative optimization procedure (IOP), a global optimization scheme similar to the one described in Ref. [3], was implemented to find an optimal solution to the SDW model parameters and quantify their uncertainties.

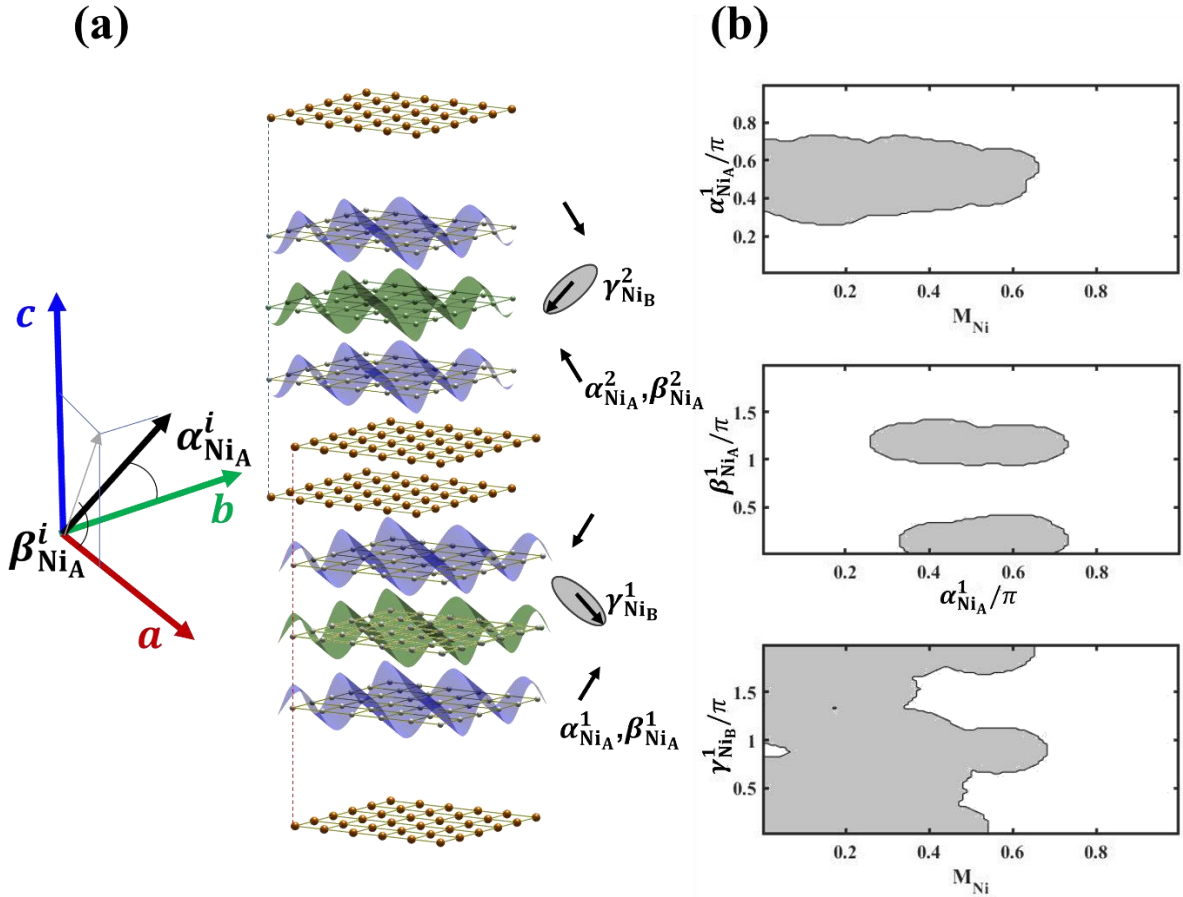

Figure S8: Modeling and data-fitting for the intermediate phases of $Pr_4Ni_3O_{10}$ and the low-temperature phase of $La_4Ni_3O_{10}$. (a) In the considered scenarios, only the Ni layers have a magnetic moment, and all the other layers are considered to be non-magnetic. The Pr/La A layers are also shown to compare with Fig. 5. The overlapping surfaces and the black arrows represent the cosine component of the SDW, $\cos(\boldsymbol{q_{SDW}} \cdot r_i)$ and the corresponding polarization, $\boldsymbol{\sigma}_i^\alpha$, respectively (see Eq. 5). Note that the wavy surfaces in Fig. 5 have a different meaning from what is shown here, in that what is plotted there is $\boldsymbol{\sigma}_i^\alpha \cdot \cos(\boldsymbol{q_{SDW}} \cdot r_i)$. A single Ni-trilayer is parameterized by three local rotation angles ($\alpha_i$, $\beta_i$, and $\gamma_i$)

and a global amplitude $M_{Ni}$, subject to the constraint that the polarization of the upper (middle) layer is equal and opposite (perpendicular) to that of the bottom layer. $M_{Ni} = M_{Ni_B}/M_{Ni_A}$ is the ratio of the middle and outer layer Ni moments. The local variables are defined as $\alpha^i_{Ni_A}$ — the polar angle of the bottom Ni layer polarization with respect to $b$, $\beta^i_{Ni_A}$ — the azimuthal angle in the a-c plane w.r.t. $a$ of the bottom Ni layer polarization, and $\gamma^i_{Ni_B}$ — the azimuthal angle of the middle Ni-layer polarization in the plane perpendicular to the outer layer polarization. (b) The 2D projections of the 4-parameter optimal region (gray contours) were found by fitting the $IntC$ data to the single tri-layer model. The calculated $I(Q)$ for the model with no moment on the inner plane ($M_{Ni} = 0$, $\alpha_{Ni_A} = \pi/2$, $\beta_{Ni_A} = 0$, i.e., outer plane moments along $a$) in the optimal region is shown in Fig. S9 (a) in comparison to the experimental data.

The optimal region learned from IOP for the SDW model with single trilayer stacking ($L_n = 1$) and the Pr$_4$Ni$_3$O$_{10}$ $IntC$ data is shown in Fig. S8b. For this particular model, the parameter space was four-dimensional ($M_{Ni}$, $\alpha^1_{Ni_A}$, $\beta^1_{Ni_A}$ and $\gamma^1_{Ni_B}$ parameters as defined in the caption of Fig. S8), and Fig. S8b shows the 2D projection of the 4D manifold of possible solutions subject to a cutoff, $\hat{\chi}^2 < C^2$. ($\hat{\chi}^2$ is the low-cost estimator of the cost function given in *Methods: Optimization*). Figure S9a shows the calculated $I(Q)$ for the particular point ($M_{Ni} = 0$, $\alpha^1_{Ni_A} = \pi/2$, $\beta^1_{Ni_A} = 0$) in the optimal region of Fig. S8b in comparison to the experiment. The relative moment of the middle Ni layer, $M_{Ni}$, is found to be less than 0.6. The zero intensity at $l = 0$ implies that either there is a node in the middle layer ($M_{Ni} = 0$) or its polarization is perpendicular to the Ni outer layer. Moreover, the middle layer polarization parameter $\gamma^1_{Ni_B}$ should be either 0 or $\pi$ ($\sigma_{Ni_B}$ along $c$) for higher values of $M_{Ni}$, and it obviously becomes unconstrained as $M_{Ni} \to 0$ (see Fig. S8b bottom panel). The uncertainty of $\alpha^1_{Ni_A}$ and $\beta^1_{Ni_A}$ is approximately $\pm 0.2\pi$ at the center of the optimal region. This single trilayer model is obviously incomplete and cannot capture the intensity profile of the experimental diffuse scattering data. The experimental peak at $l = 2$ is narrower compared to the ($L_n = 1$) model indicating coherent stacking over a longer distance is required. Thus, IOP was also run for the SDW models with double ($L_n = 2$) and triple ($L_n = 3$) trilayer stacking. After exploring both seven and ten-dimensional parameters spaces $\left(M_{Ni}, \{\alpha^i_{Ni_A}, \beta^i_{Ni_A}, \gamma^i_{Ni_B}\}_{i=1...L_{TL}}\right)$, respectively, the optimal regions are found to be in good agreement with each other. The relations: $\alpha^1_{Ni_A} = \alpha^2_{Ni_A} = \cdots$, $\beta^1_{Ni_A} = \beta^2_{Ni_A} = \cdots$ were found with comparable uncertainties to the ($L_n = 1$) model. The calculated $I(Q)$ for special points where $\alpha^1_{Ni_A} = \alpha^2_{Ni_A} = \cdots = \pi/2$ and $\beta^1_{Ni_A} = \beta^2_{Ni_A} = \cdots = 0$ in the optimal regions are shown in Fig. S9b and S9c for the $L_n = 2$ and $L_n = 3$ models, respectively.

Careful observation of the effect of each model on the fit to the $l = 2$ peak reveals that each model captures the peak profile in different parts of reciprocal space, indicating that a more sophisticated model is required. SDW models allowing for multiple stacking lengths $L_n$ simultaneously are considered with the constraints of $M_{Ni} = 0$, $\{\alpha^i_{Ni_A} = \frac{\pi}{2}, \beta^i_{Ni_A} = 0\}_{i=1...L_n}$ and the probability of each stacking of length $\rho(L_n)$ is refined. The best solution for the SDW model with $L_n \leq 2$ is shown in Fig. S9d. In contrast to the previous models, the $L_n \leq 2$ model fits the experimental data markedly better. Furthermore, a more complex model of $L_n \leq 16$ was tried with La$_4$Ni$_3$O$_{10}$ low-$T$ data and Pr$_4$Ni$_3$O$_{10}$ $IntC$ data, and the best fits are shown in Figs. 4a and 4b, respectively. The corresponding $\rho(L_n)$ are shown in Fig. S11a and S11b. The fitting of $\rho(L_n)$ allowed for optimization of the observed peak shapes, which were complex and contained both broad (captured by low $n$) and narrow components (captured by high $n$). The best fit, as shown in Fig. 4a for La$_4$Ni$_3$O$_{10}$, consisted of a rapidly decaying $\rho(L_n)$ with increasing $n$ – as would be expected for quasi-2D order – combined with a nonzero minority contribution captured by a component with larger $n$ which breaks the trend of decay. We suggest an inhomogenous picture in which the majority of the order is short-ranged (low $n$) but that a minority long-range ordered component exists. As demonstrated in Fig. S12, the long-range components $L_n = 7, 12, 13$ are needed to capture sharper features (peaks) that coexist with a broad signal, similar to the Pr$_4$Ni$_3$O$_{10}$ $IntC$ data.

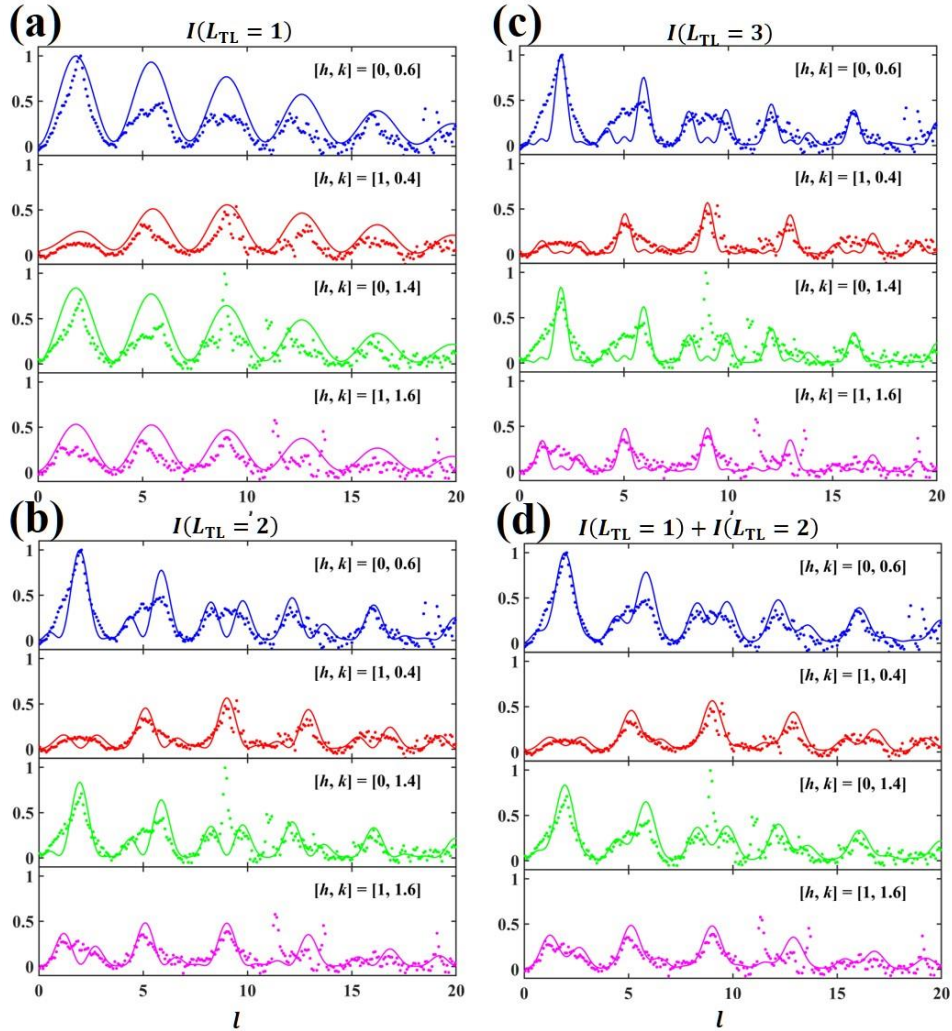

Figure S9: The calculated structure factors $I(\mathbf{Q})$ for the optimized models for four different scenarios. Models of stacking length ($L_n$) (a) of a single trilayer, (b) of a double trilayer, (c) of three trilayers, and (d) a sum of single and double trilayers. For (a), (b), and (c), the M$_{\text{Ni}}$, $\{\alpha, \beta, \gamma\}$ parameters are refined, and the optimal regions are found to be similar. Given this, the M$_{\text{Ni}}$, $\{\alpha, \beta\}$ parameters are then fixed to $0, \{\frac{\pi}{2}, 0\}$

and the probability distribution as a function of $L_n$ was refined instead for the models having multiple stacking lengths (e.g., (d)). The refined probabilities are $\rho(L_n = 1) = 0.36$ and $\rho(L_n = 2) = 0.64$.

The Pr$_4$Ni$_3$O$_{10}$ diffuse peak profiles in the intermediate phase while warming ($IntW$) data are different from those of $IntC$. Some of the "pointy" peaks in $IntC$ have become broad and squared up, and additional structure has appeared in other smooth regions. We found the relations $M_{Ni} < 0.6$, $\alpha_{Ni_A}^{4n} = \alpha_{Ni_A}^{4n+1} = -\alpha_{Ni_A}^{4n+2} = -\alpha_{Ni_A}^{4n+3} = \pi/2$ and $\beta_{Ni_A}^{4n} = \beta_{Ni_A}^{4n+1} = \beta_{Ni_A}^{4n+2} = \beta_{Ni_A}^{4n+3} = \cdots = 0$ ($n = 1, \ldots, L_n/4$) from optimization attempts on the $IntW$ data with single stacking length SDW models. A more complex SDW model for $L_n \leq 16$ was fit to the $IntW$ data with $M_{Ni} = 0$ and the aforementioned relations for $\{\alpha, \beta\}$. The $I(\mathbf{Q})$ for the best solution is shown in Fig. 4c, and the refined $\rho(L_n)$ is shown in Fig. S11c. $\rho_{IntW}(L_n)$ shows a rapid fall-off as a function of $L_n$ similar to $IntC$ along with longer-range stacking. Note that the two SDW models used for $IntC$ and $IntW$ cases are equivalent to each other for $L_n \leq 2$.

The fits to the equivalent peaks arising from twins (shown in Fig. 4) are also shown in Fig. S10, and we find that the refined SDW models are comparable as expected.

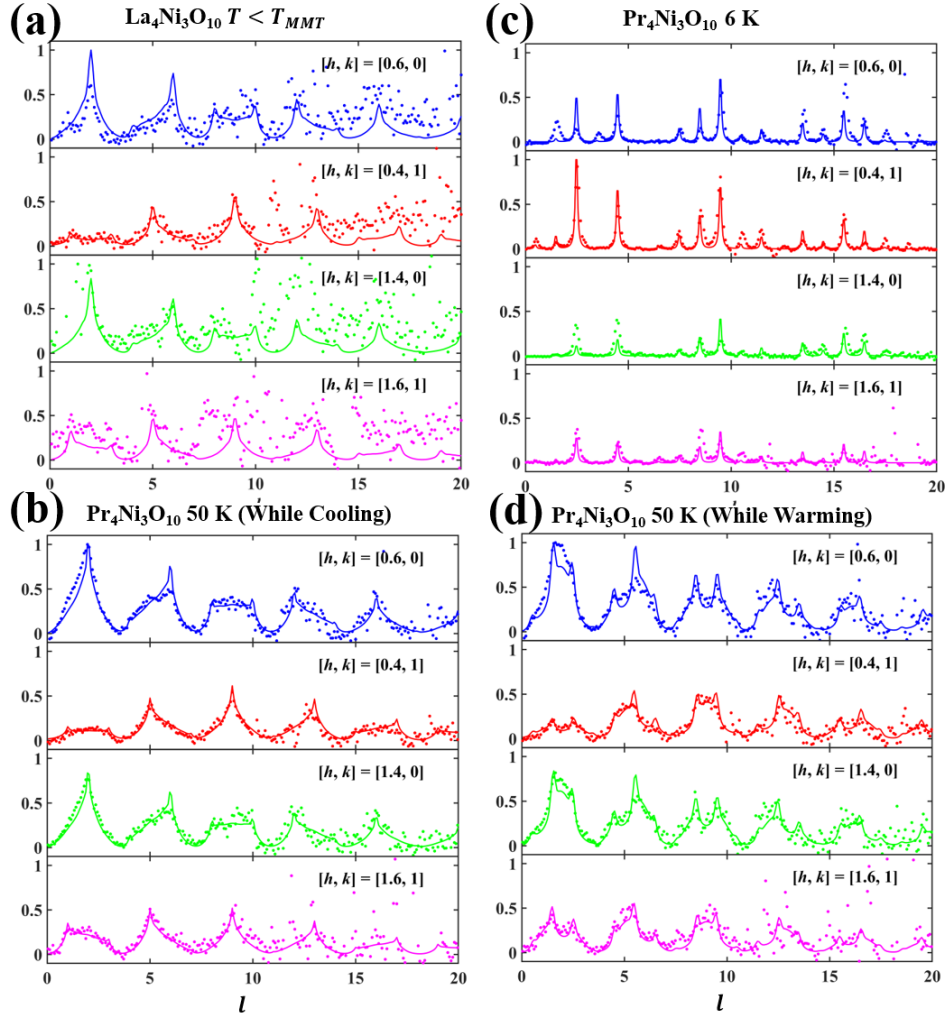

Figure S10: Comparison of neutron data and the corresponding optimized magnetic models. (a) La$_4$Ni$_3$O$_{10}$ data collected at $T < T_{MMT}$, (b) Pr$_4$Ni$_3$O$_{10}$ data at 50 K measured while cooling from 200 K, (c) Pr$_4$Ni$_3$O$_{10}$ data collected at base-temperature ($\approx$6 K) and (d) Pr$_4$Ni$_3$O$_{10}$ data measured at 50 K while warming from base-temperature. Because of the temperature independent nature of the structure factors below $T_{MMT}$ for La$_4$Ni$_3$O$_{10}$ (see Fig. 3d), all the low-temperature data were combined in panel (a). Each panel shows line cuts of 3D scattering data along $l$ for four different combinations of $h \pm 0.1$ and $k \pm 0.1$: (0.6, 0), (0.4, 1), (1.4, 0) and (1.6, 1).

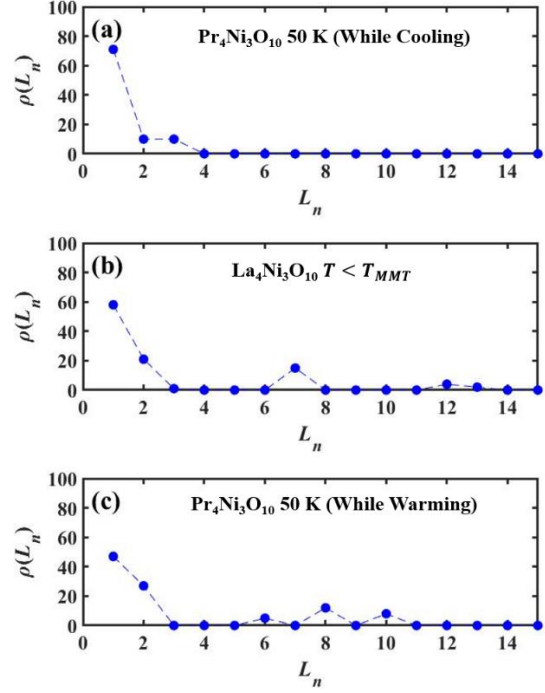

Figure S11: The probability distributions $\rho(L_n)$ as a function of the stacking length ($L_n$) obtained by fitting neutron diffuse scattering data sets. The extracted $\rho(L_n)$ for (a) the $Pr_4Ni_3O_{10}$ data set measured at 50 K after cooling from higher temperatures, (b) the $La_4Ni_3O_{10}$ data set measured below $T_{MMT} \approx 148$ K and (c) the $Pr_4Ni_3O_{10}$ data set measured at 50 K after warming from the base temperature are shown.

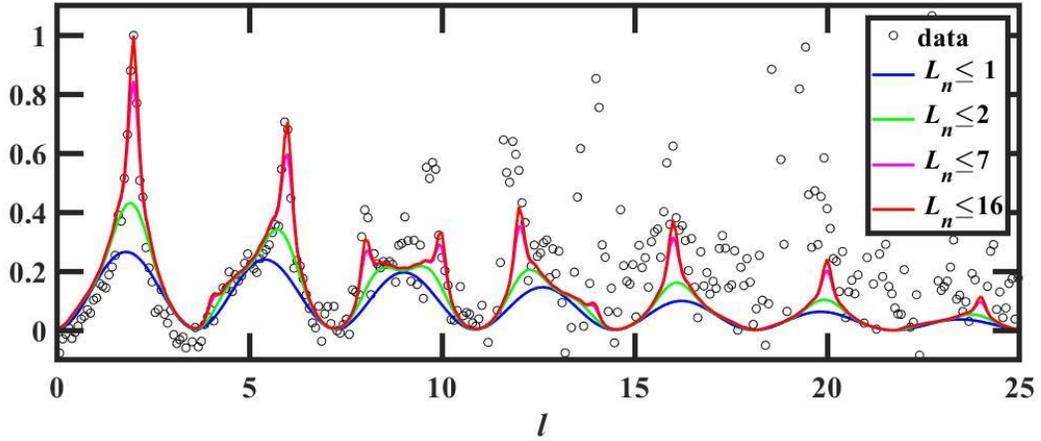

Figure S12: The calculated structure factor $I(\mathbf{Q})$ of the composite SDW model with maximum stacking lengths ($L_n$) up to $n = 1, 2, 7$ and $16$ for $La_4Ni_3O_{10}$.

## IV. Modeling and optimization of the low-temperature phase

It is evident from the temperature dependence of $I(\mathbf{Q})$ along $l$ (see Fig. S1a and Figure S5) that the $IntW$ phase gains spectral weight at the magnetic Bragg peak positions ($l = n/2$, where $n$ is odd) of the $LowT$ phase by suppressing intensities at diffuse peak centers found in the $IntC$ pattern. Thus, that the same local SDW stacking pattern is shared by both $LowT$ and $IntW$ phases is a reasonable hypothesis.

As shown in Figure S13a and S13b, $I(\mathbf{Q})$ was calculated for two simplified models (Ni layers only), varying the direction of polarization, with the $IntW$ stacking pattern enforced. The Ni-only models have many extra peaks, some of which have stronger intensities. Thus, we implemented a complex model including all the Pr layers and parameterized with 25 variables as described in *Methods: Density wave models and simulations*. From IOP, we found that $M_{Pr_A} \approx 1$, $M_{Pr_B} \approx 0$, $M_{Ni} < 0.5$ and $\alpha_{Pr_A} = \pi$, $\beta_{Pr_A} = 0$, $\alpha_{Ni_A}^2 = \beta_{Ni_A}^1 = \beta_{Ni_A}^2 = 0$ for $\alpha_{Ni_A}^1 = \pi$ and the best solution is shown in Fig. 4c. Note that here we fit both neutron data and RXS data simultaneously. The RXS data favor Pr polarizations to be along $\mathbf{q}_{SDW}$. This model reproduces most of the magnetic peaks and relative intensities except for a few peaks, including (0.6, 0, 1.5) and (0.6, 0, 3.5). More importantly, this model does not introduce extra peaks where no magnetic signal is observed experimentally. Adding emphasis on the (0.6, 0, 1.5) peak favors the models where only the polarization of the Ni layers are canted away from $\mathbf{q}_{SDW}$ ($\alpha_{Ni_A}^2 = \delta, \alpha_{Ni_A}^1 = \pi - \delta$) while keeping the Pr polarization $\parallel \mathbf{q}_{SDW}$. As shown in Figure S13c, the model with $\alpha_{Ni_A}^2 = \pi/4, \alpha_{Ni_A}^1 = \pi - \pi/4$ recovers the (0, 0.6, 1.5) peak at the cost of extra peaks relative to experiment.

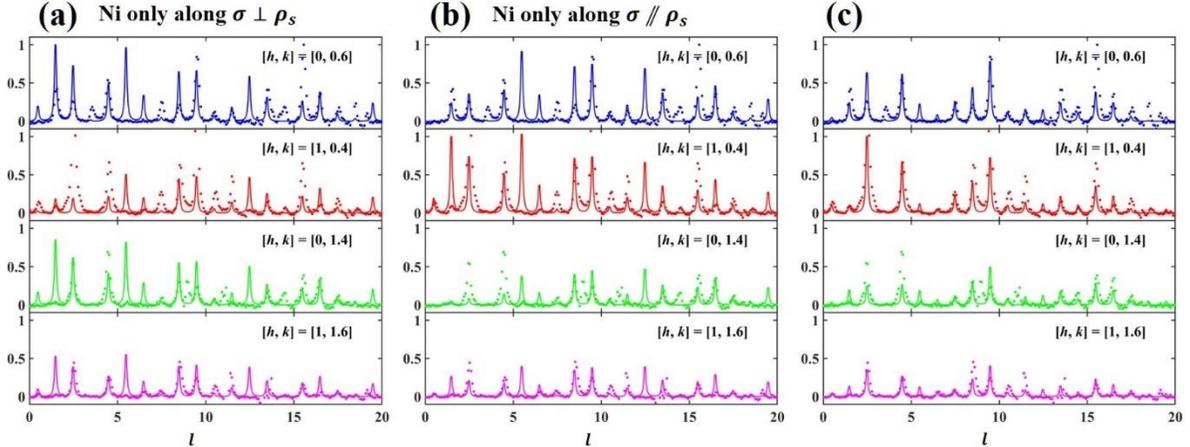

Figure S13: Comparison of the calculated $I(\mathbf{Q})$ for different long-range ordered magnetic models and experimental data collected on $Pr_4Ni_3O_{10}$ at 6 K. The model considered in (a) is the same as depicted in Fig. 5c. It is similar for panel (b) as well, except the SDW polarization is along the SDW propagation vector $\mathbf{q}_{SDW}$ rather than being perpendicular. The model considered in panel (c) is similar to Fig. 5b, except the Ni-layer polarizations are globally rotated by $\pi/4$ relative to Pr. This model can recover the peak intensity at (0.6, 0, 1.5) at the cost of extra peaks appearing.

## V. Phenomenological model and analysis

Figure 5 depicts the best SDW models extracted from the data-fitting for each case. Briefly, the outer layer Ni polarization orients perpendicular to $q_{SDW}$ in the intermediate temperature phases, while the polarization of both the outer layer Ni and the Pr A-site layers align parallel to $q_{SDW}$ in the low-$T$ phase. The local stacking pattern of Ni layers also switches from [1, 0, -1], [1, 0, -1], [1, 0, -1], [1, 0, -1] to [1, 0, -1], [1, 0, -1], [-1, 0, 1], [-1, 0, 1] below $T_f$ in the cooling process from the intermediate-$T$ phase to the low-$T$ phase, and it stays the same even at intermediate temperatures during the warming-up process, until it is completely erased above $T_{MMT}$. Note that only the stacking pattern stays the same during the warming-up process, but the polarization returns to aligning perpendicular to $q_{SDW}$ (see Fig. 5 or S15). We devised an effective Hamiltonian to understand this behavior and studied its properties.

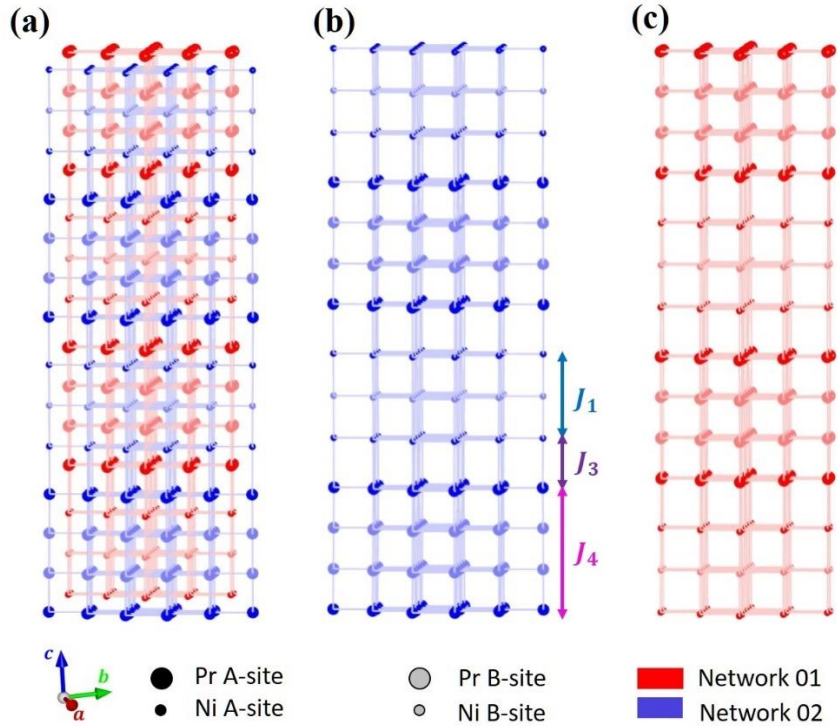

Figure S14: Illustration of dissecting the magnetic unit-cell of the $Pr_4Ni_3O_{10}$ *LowT* phase into two equivalent orthorhombic lattices. (a) The complete magnetic unit cell with Ni and Pr ions. (b) and (c) are the two sublattices related by the face centering operation. The inter-sublattice exchanges considered in the proposed effective Hamiltonian are also shown. $J_2$ is the only intra-sublattice exchange as shown in Fig. 1b.

Before going into detail about the phenomenological model, we discuss a few facts about the nuclear structure. The nuclear unit cell can be dissected into two orthorhombic sublattices related by the face centering translation, as shown in Figure S12. The Ni trilayers and the $Pr_2O_2$ blocks are alternatively stacked in each sublattice. The proposed exchange interactions between layers and the ligand environment are shown in Fig. 1b. Note the exchange interactions are defined in detail between the polarization vectors of magnetic layers as formulated in Eq. 1 of the main text.

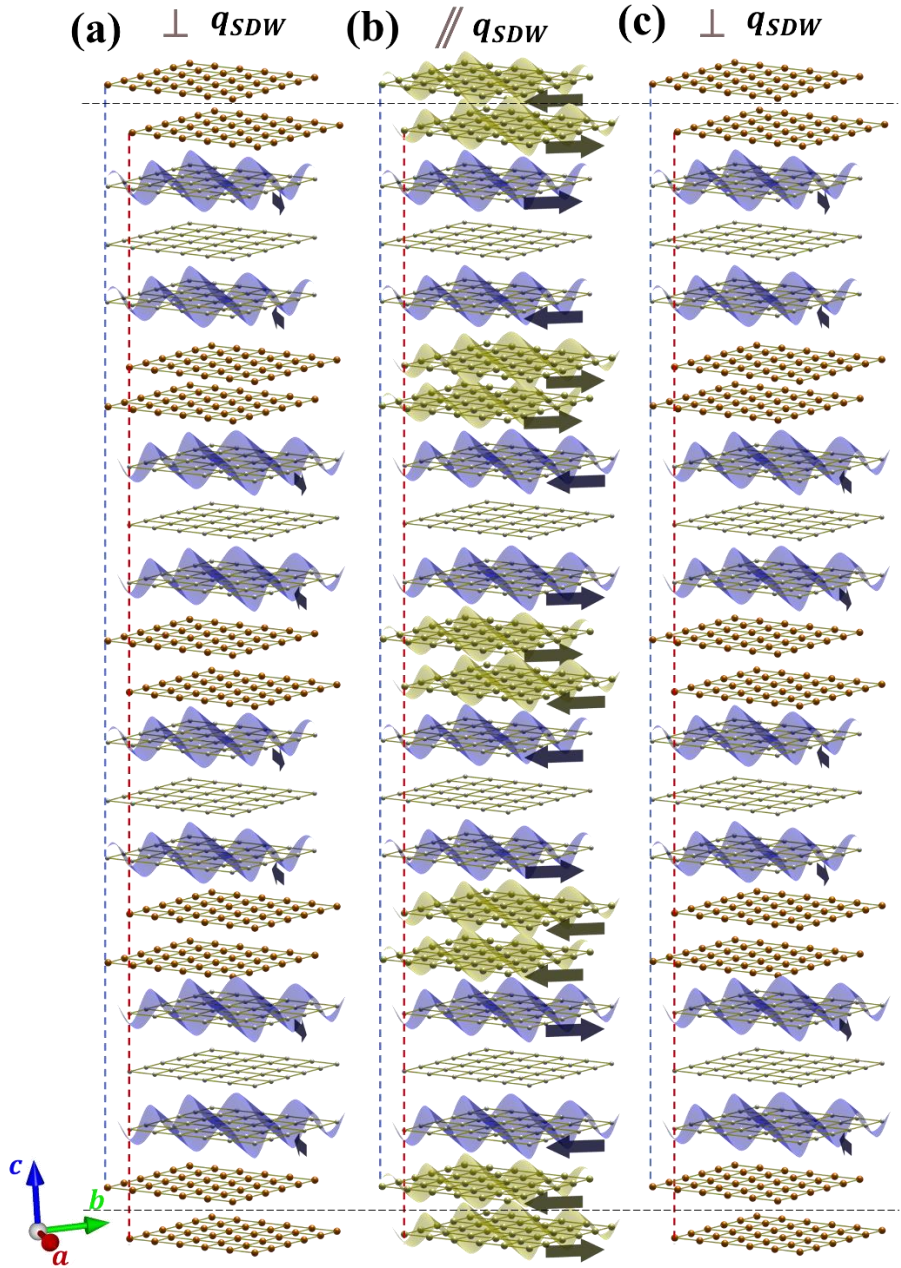

Figure S15: The optimized magnetic models for (a) the intermediate-temperature phase ($T_f < T < T_{MMT}$) while cooling from the paramagnetic phase, $IntC$, (b) the low-temperature phase ($T < T_f$), $LowT$ and (c) the intermediate-temperature phase ($T_f < T < T_{MMT}$) while warming from the low-T phase, $IntW$. The corresponding SDWs are plotted in the orthorhombic $\boldsymbol{ab}$ plane. Only Ni A and Pr A layers are shown. Note that the SDWs shown here are dissected into their sinusoidal components (wave-forms) and polarizations (arrows). In contrast, Fig. 5 in the main text and Fig. 4 of Ref. [2] depicts SDWs convoluted by the sign of the polarizations.

Note that $D$ and $M_{Pr_A}$ are the strength of the single ionic anisotropy (SIA) with the easy-axis direction as $\hat{b}$, and the induced moment of the Pr A site, respectively. $M_{Pr_A}$ is a function of temperature as observed in the RXS experiments (see Fig. S7b). This effective Hamiltonian was deduced from a symmetry analysis of the nuclear structure of $Pr_4Ni_3O_{10}$ (space group: *Bmab*) using the PSI-SpinW package [4]. Since we are modeling only the diffraction data in this work, these exchange parameters are ill-posed. However, we can find the following limits where the SDW models can be energetically stabilized.

$$\begin{gathered} J_a > J_b, J_c \geq 0 \\ J_d > 0 \\ J_e > J_f, J_g \geq 0 \\ J_h < 0 \\ 2|J_d| < |J_h| \\ D > 0 \end{gathered} \tag{S2}$$

The three SDW models ($IntC$, $LowT$, and $IntW$) are connected through a continuous transformation in configurational space as parameterized by the global rotation angle ($\Theta$) and the canting angle ($\varphi$) of one nuclear unit cell relative to the other (see Fig. S16). For example, the $IntC$, $LowT$, and $IntW$ models are located at $[\varphi, \Theta]$ =[0, 0], [$\pi$, $\pi/2$], and [$\pi$, 0], respectively.

Figure S16 depicts the configurations as a function of $\varphi$ at $\Theta = 0$. The polarization vector of each layer is also marked on the side, and two sets of labels are used. One ($\sigma_j$) for those fixed and the other ($\sigma_{jv}$) for those rotated by $\varphi$. Each polarization vector can be written in $[\varphi, \Theta]$ as,

$$\begin{gathered} \sigma_2 = \sigma_4 = \sigma_6 = \sigma_8 = -\sigma_1 = -\sigma_3 = -\sigma_5 = -\sigma_7 \equiv [\sin(\Theta), \quad \cos(\Theta), \quad 0] \\ \sigma_{2v} = \sigma_{5v} = \sigma_{6v} = \sigma_{8v} = -\sigma_{1v} = -\sigma_{3v} = -\sigma_{4v} = -\sigma_{7v} \\ \equiv [\sin(\varphi + \Theta), \quad \cos(\varphi + \Theta), \quad 0] \end{gathered} \tag{S3}$$

Eq. 1 of the main text can be re-written for the magnetic unit-cell with periodic boundary conditions:

$$\begin{gathered} E \propto J_1[\sigma_2 \cdot \sigma_3 + \sigma_{2v} \cdot \sigma_{3v} + \sigma_{6v} \cdot \sigma_{7v} + \sigma_6 \cdot \sigma_7] + J_2[\sigma_2 \cdot \sigma_7 + \sigma_3 \cdot \sigma_{2v} + \sigma_{3v} \cdot \sigma_{6v} + \sigma_{7v} \cdot \sigma_6] + J_3 \cdot M_{Pr_A}[\sigma_1 \cdot \sigma_2 + \sigma_3 \cdot \sigma_4 + \sigma_{1v} \cdot \sigma_{2v} + \sigma_{3v} \cdot \sigma_{5v} + \sigma_{4v} \cdot \sigma_{6v} + \sigma_{7v} \cdot \sigma_{8v} + \sigma_5 \cdot \sigma_6 + \sigma_7 \cdot \sigma_8] + J_4 \cdot M_{Pr_A}^2[\sigma_{1v} \cdot \sigma_8 + \sigma_4 \cdot \sigma_{4v} + \sigma_{5v} \cdot \sigma_5 + \sigma_{8v} \cdot \sigma_1] - D \cdot \\ M_{Pr_A}^2[(\sigma_1 \cdot \hat{x})^2 + (\sigma_{1v} \cdot \hat{x})^2 + (\sigma_4 \cdot \hat{x})^2 + (\sigma_{4v} \cdot \hat{x})^2 + (\sigma_{5v} \cdot \hat{x})^2 + (\sigma_5 \cdot \hat{x})^2 + (\sigma_{8v} \cdot \hat{x})^2 + (\sigma_8 \cdot \hat{x})^2] \end{gathered} \tag{S4}$$

and then further refined when including the $J$ anisotropies:

$$
\begin{aligned}
E \propto\ & J_a[\sigma_2^x\sigma_3^x + \sigma_{2v}^x\sigma_{3v}^x + \sigma_{6v}^x\sigma_{7v}^x + \sigma_6^x\sigma_7^x] + J_b[\sigma_2^y\sigma_3^y + \sigma_{2v}^y\sigma_{3v}^y + \sigma_{6v}^y\sigma_{7v}^y + \sigma_6^y\sigma_7^y] + \\
& J_c[\sigma_2^z\sigma_3^z + \sigma_{2v}^z\sigma_{3v}^z + \sigma_{6v}^z\sigma_{7v}^z + \sigma_6^z\sigma_7^z] + J_d[\sigma_2\cdot\sigma_7 + \sigma_3\cdot\sigma_{2v} + \sigma_{3v}\cdot\sigma_{6v} + \sigma_{7v}\cdot\sigma_6] + \\
& J_e\cdot M_{Pr_A}[\sigma_1^x\sigma_2^x + \sigma_3^x\sigma_4^x + \sigma_{1v}^x\sigma_{2v}^x + \sigma_{3v}^x\sigma_{5v}^x + \sigma_{4v}^x\sigma_{6v}^x + \sigma_{7v}^x\sigma_{8v}^x + \sigma_5^x\sigma_6^x + \sigma_7^x\sigma_8^x] + \\
& J_f\cdot M_{Pr_A}[\sigma_1^y\sigma_2^y + \sigma_3^y\sigma_4^y + \sigma_{1v}^y\sigma_{2v}^y + \sigma_{3v}^y\sigma_{5v}^y + \sigma_{4v}^y\sigma_{6v}^y + \sigma_{7v}^y\sigma_{8v}^y + \sigma_5^y\sigma_6^y + \sigma_7^y\sigma_8^y] + \\
& J_g\cdot M_{Pr_A}[\sigma_1^z\sigma_2^z + \sigma_3^z\sigma_4^z + \sigma_{1v}^z\sigma_{2v}^z + \sigma_{3v}^z\sigma_{5v}^z + \sigma_{4v}^z\sigma_{6v}^z + \sigma_{7v}^z\sigma_{8v}^z + \sigma_5^z\sigma_6^z + \sigma_7^z\sigma_8^z] + \\
& J_h\cdot M_{Pr_A}^2[\sigma_{1v}\cdot\sigma_8 + \sigma_4\cdot\sigma_{4v} + \sigma_{5v}\cdot\sigma_5 + \sigma_{8v}\cdot\sigma_1] \\
& -D\cdot M_{Pr_A}^2[(\sigma_1\cdot\hat{x})^2 + (\sigma_{1v}\cdot\hat{x})^2 + (\sigma_4\cdot\hat{x})^2 + (\sigma_{4v}\cdot\hat{x})^2 + (\sigma_{5v}\cdot\hat{x})^2 + (\sigma_5\cdot\hat{x})^2 \\
& \quad + (\sigma_{8v}\cdot\hat{x})^2 + (\sigma_8\cdot\hat{x})^2]
\end{aligned}
\quad\text{(S5)}
$$

By substituting the polarization vectors from Eq. S3, the expression simplifies to Eq. 2 of the main text.

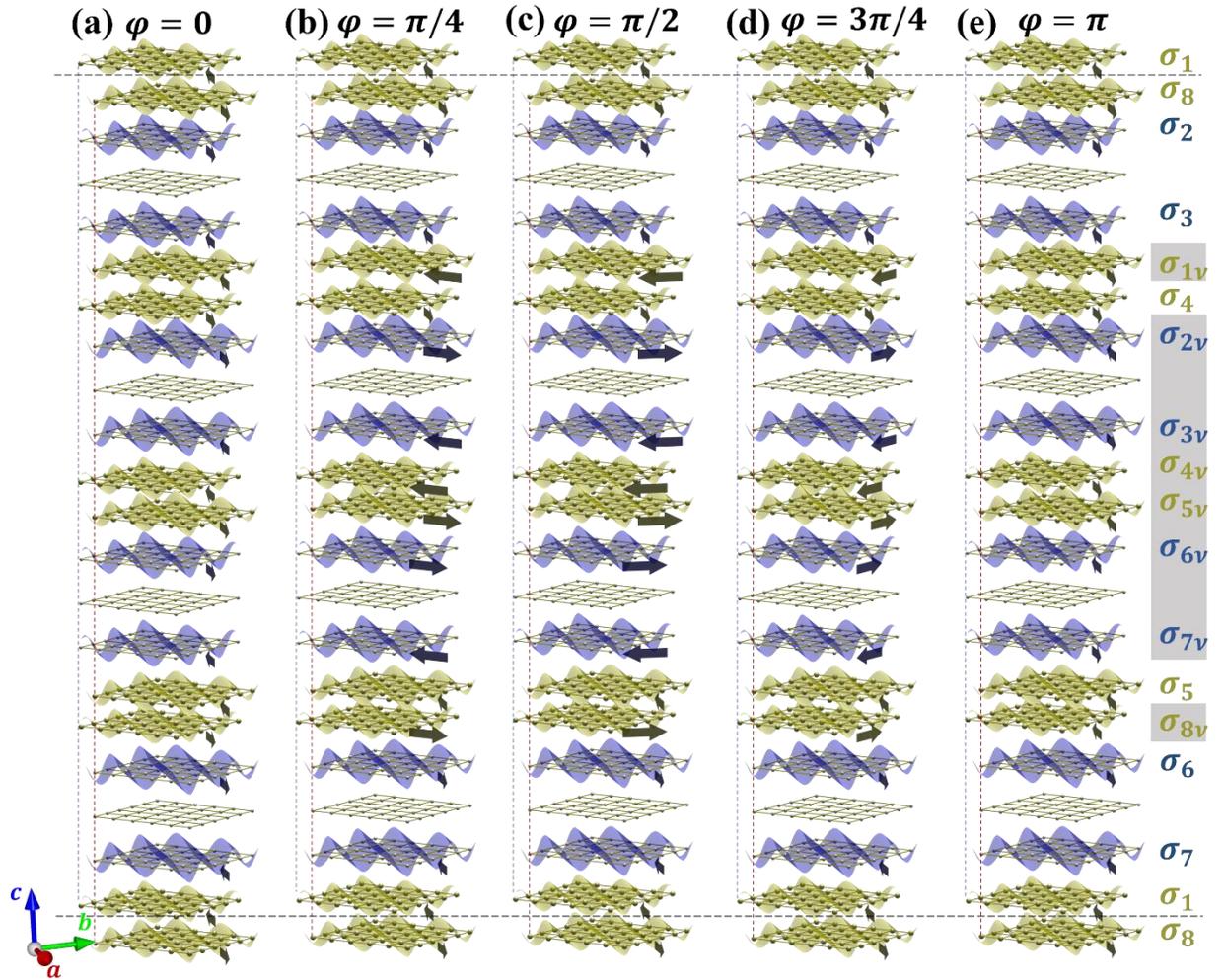

Figure S16: The magnetic models along a line parameterized by the in-plane canting angle $\varphi$ connecting $IntC$ ($\varphi = 0$) and $IntW$ ($\varphi = \pi$) models in the configurational space. $\Theta = 0$ for each. Note that the SDWs

shown here are decomposed into their sinusoidal components (wave-forms) and polarizations (arrows) similarly to Fig. S15. The gray color boxes highlight the in-plane twist operation (twist angle $\varphi$) on the polarization vectors with respect to the rest of the layers.

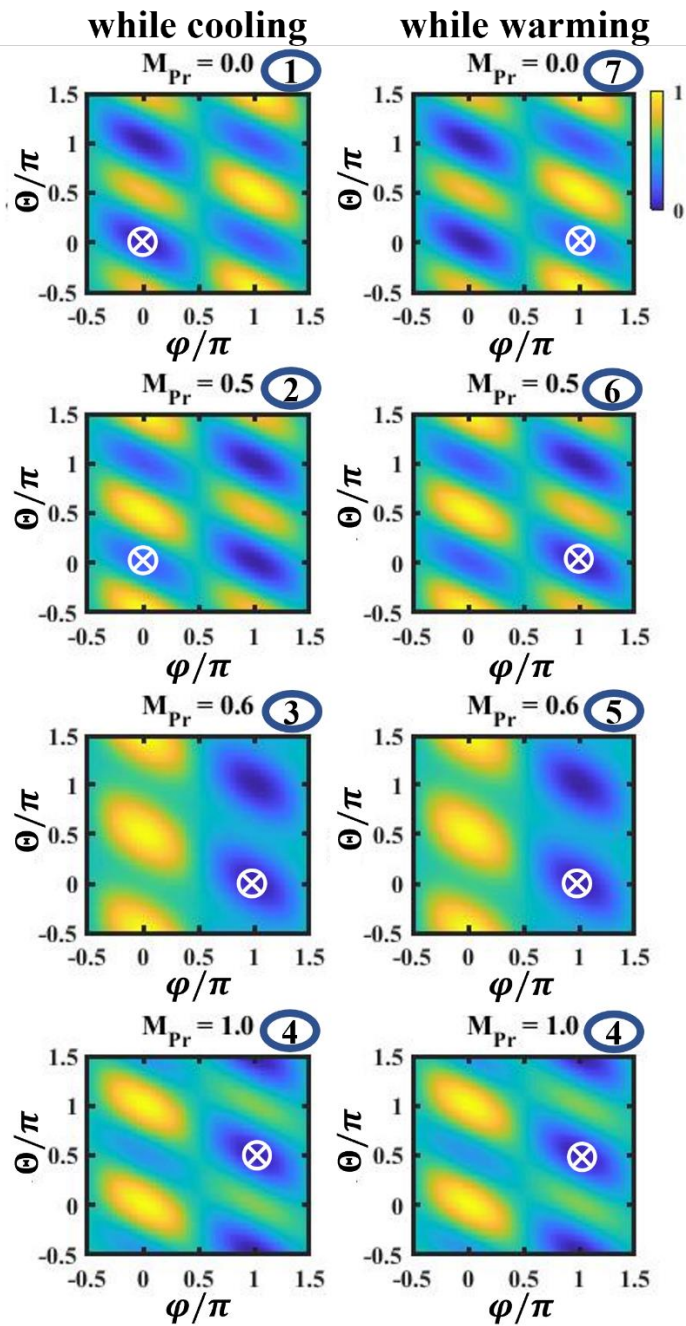

Figure S17: The evolution of the 2D energy landscape $E(\varphi, \Theta)$ with $M_{Pr}$ for the exchange parameters $J_a = 1, \frac{J_b}{J_a} = 0.5, \frac{J_c}{J_a} = 0, \frac{J_d}{J_a} = 0.1, \frac{J_e}{J_a} = 0.5, \frac{J_f}{J_a} = 0.25, \frac{J_g}{J_a} = 0, \frac{J_h}{J_a} = -0.3$ and $\frac{D}{J_a} = 0.5$. The order of $E(\varphi, \Theta)$ under the process of cooling from $T > T_f$ to $T_{base}$ and warming back to $T > T_f$ is marked by the labels 1-7. The white color markers are a guide to show the stable configuration of the system at a particular step in the process. Note that it is globally stable during the cooling process (left-hand side), and it is kinetically stable during the warming process (right-hand side). Axes are scaled by π.

Eq. 2 of the main text was used to generate Figures 6, S17, and S18. Both Figure 6 and S17 are generated for a selected parameter set $(J_a = 1, \frac{J_b}{J_a} = 0.5, \frac{J_c}{J_a} = 0, \frac{J_d}{J_a} = 0.1, \frac{J_e}{J_a} = 0.5, \frac{J_f}{J_a} = 0.25, \frac{J_g}{J_a} = 0, \frac{J_h}{J_a} = -0.3$ and $\frac{D}{J_a} = 0.5)$ by increasing $M_{Pr}$ from 0 to 1 and then decreasing it back to 0, signifying the experimental process. Note that $M_{Pr_A} = 0$ in the intermediate-$T$ phase and $M_{Pr_A} \approx 1$ at the base temperature.

For $M_{Pr_A} = 0$, there is a global minimum at $[\varphi, \Theta] = [0,0]$ and a local minimum at $[\pi, 0]$ separated by an energy barrier, $E_b$. Cooling the system from a high temperature above $T_{MMT}$ would naturally place the system in the global minimum where the $IntC$ stacking stabilizes. Upon further cooling, $M_{Pr_A}$ increases and the energy landscape evolves drastically such that the global minimum is no longer at $[0,0]$ (when $M_{Pr_A} = 0.5$), and the minimum energy SDW stacking will be rearranged. At $M_{Pr_A} = 1$ ($T = T_{base}$), the global minimum is at $[\pi, \pi/2]$, and the corresponding stacking pattern is as shown in Fig. 5b and S15b. During the warming process (decreasing $M_{Pr_A}$), the system will be pushed towards $[\pi, 0]$ rather than its initial state $[0,0]$ and will get trapped by local energy barriers. As this metastable state is protected by the energy barrier $E_b$, the system will remain trapped even for $M_{Pr_A} = 0$ until the thermal energy of the system becomes high enough to overcome $E_b$. Thus, the system remembers the stacking pattern of the $LowT$ phase at intermediate temperatures while warming, causing the temperature-hysteresis observed in neutron diffraction. Importantly, this behavior is robust for any parameter set within the range given in Eq. S2.

A sensitivity analysis of Eq. 2 (main text) for the exchange parameters was done. Figure S18 shows the evolution of the energy landscape for five different exchange parameter sets listed in Table S2.

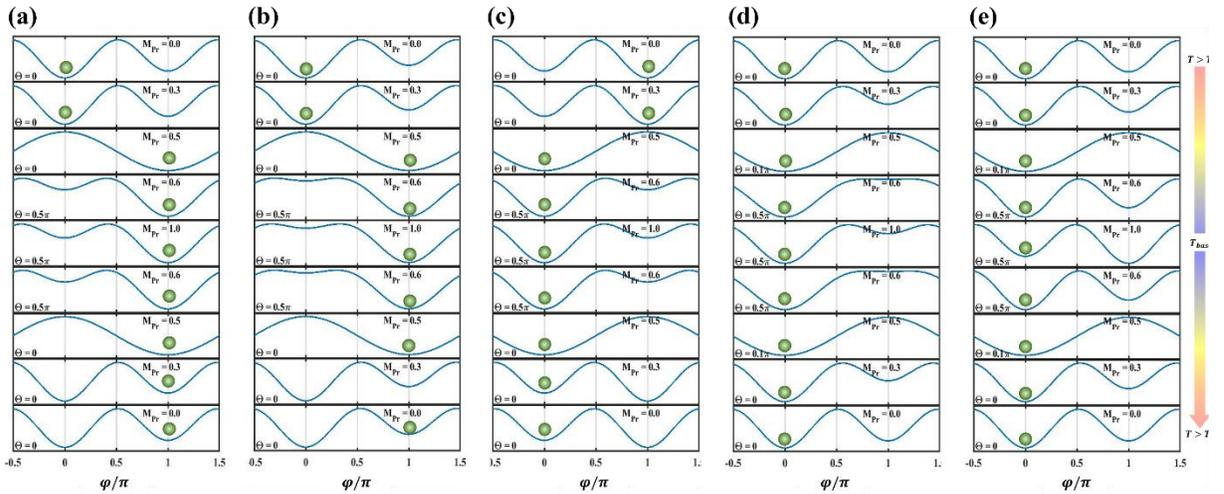

Figure S18: Evolution of the energy landscape $E(\varphi)$ with $M_{Pr_A}$ for five different sets of exchange parameters as given in Table S2. Note that the global rotation parameter $\Theta$ is optimized for each panel. The subpanel labels from (a) to (e) match the parameter set indices from 1 to 5 in Table S2. The green color markers are a guide to show the stable configuration of the system at a particular step in the process.

The $J_1$ and $J_3$ exchanges are considered to be anisotropic, as defined in Eq. 1 (main text). The corresponding exchange pathways are $Ni_A$-O-$Ni_B$-O-$Ni_A$ and $Ni_A$-O-$Pr_A$, respectively. All M-O-M bond angles are approximately 168° (see Table S3) and canted along $\hat{b}$. Thus, one might naively assume that both exchanges are anisotropic with similar structures. To stabilize the intermediate phase structures in which all the polarization vectors are perpendicular to $q_{SDW}$ ($\| \hat{b}$), the condition $J_a > J_b, J_c \geq 0$ should be satisfied. Having a similar condition for $J_3$ will not allow polarization vectors to align parallel to $q_{SDW}$ in the low-temperature phase. Thus, an additional single ion anisotropy (SIA) term, $D$, for $Pr_A$ sites is necessary to stabilize the *LowT* model; this can be justified by its distinctly anisotropic ligand environment.

Table S2: Exchange parameter sets used to study the evolution of the energy landscape $E(\varphi)$ with $M_{Pr}$. The definitions of each exchange parameter can be found in Eq. 1 of the main text.

| # | $J_b/J_a$ | $J_c/J_a$ | $J_d/J_a$ | $J_e/J_a$ | $J_f/J_a$ | $J_g/J_a$ | $J_h/J_a$ | $D/J_a$ |
|---|---|---|---|---|---|---|---|---|
| 1 | 0 | 0 | 0.1 | 0 | 1 | 0 | -0.3 | 0 |
| 2 | 0.5 | 0 | 0.1 | 0.5 | 1 | 0 | -0.3 | 0 |
| 3 | 0 | 0 | -0.1 | 0 | 1 | 0 | 0.3 | 0 |
| 4 | 0 | 0 | 0.1 | 0 | 1 | 0 | 0.3 | 0 |
| 5 | 0 | 0 | 0.1 | 0 | 1 | 0 | -0.1 | 0 |

Alternatively, the same behavior can be seen for $D = 0$ with the condition $J_f > J_e, J_g \geq 0$ as shown in Fig. S18a and S18b. The condition $J_d > 0$ must be satisfied for the *IntC* stacking structure to have the lowest energy for $M_{Pr_A} = 0$ (the contrary is shown in Fig. S18c). Moreover, the conditions $J_h < 0$ and $2|J_d| < |J_h|$ should be satisfied to stabilize the stacking pattern of the *LowT* model and to show the memory effect/thermal hysteresis, respectively (the contrary is shown in Fig. S18d and S18e).

Table S3: Bond angles along $c$.

| Path | Angle |
|---|---|
| $Ni_A$-O-$Ni_B$ | 166.2° |
| O-$Ni_B$-O | 180° |
| $Ni_B$-O-$Ni_A$ | 166.2° |
| $Ni_A$-O-$Pr_A$ | 170° |

# VI.  Pr Crystal Electric Fields and Induced Magnetism

$Pr^{3+}$ ($4f^2$; $L=5$, $S=1$, $J=4$) is a non-Kramer's ion that, in the structure of $Pr_4Ni_3O_{10}$, occupies crystallographic sites whose point group symmetry is spanned only by one-dimensional irreducible representations. Thus, all nine crystal field levels of the $Pr^{3+}$ ions should be singlets, precluding long-range magnetic order in the absence of a magnetic field that can admix higher-lying states into the ground state. In the case of $Pr_4Ni_3O_{10}$, this magnetic field arises internally from exchange interactions with the ordered Ni sublattice. Such 'singlet magnetism' in rare earth compounds has been discussed by Bleaney [5], Trammel [6], Cooper [7] and others. As discussed in the main text and following the arguments of Rout *et al.* [8] we consider Pr B to be nonmagnetic, a conclusion supported by refinement of the neutron data presented in the main text. Thus, Pr A carries the induced moment in $Pr_4Ni_3O_{10}$. A similar induced magnetism effect has been proposed as the origin of magnetic order on the Pr sublattice of the related $n=1$ Ruddlesden-Popper nickelate, $Pr_2NiO_4$ [9,10].

With the caveat that quantitative results of such calculations should be viewed with caution, the results of a simple point-charge model of the crystal electric field (CEF) energy spectrum for both sites are shown in Fig. S19a [11]. Atomic coordinates were taken from the single crystal x-ray diffraction study of Ref. [12] and so some uncertainty in the O atom positions coordinating Pr is to be expected. Confirming the symmetry analysis above, nine singlets are found for each site. The first excited state of each ion lies $\approx$ 6 meV and 4 meV above the ground state for Pr A and Pr B, respectively, while subsequent states are found at >14 meV and are not expected to contribute significantly to the low temperature magnetic or thermodynamic behavior. For comparison, in $Pr_2NiO_4$, via refinement of inelastic neutron scattering data, Boothroyd *et. al.* report two singlets separated by 4.3 meV [10], and Rosenkranz *et al.* [13] report 6.4 meV for $PrNiO_3$. More sophisticated approaches, e.g., fitting to inelastic neutron spectra, could be helpful to better understand the crystal field levels and magnetism in $Pr_4Ni_3O_{10}$.

We measured the magnetic heat capacity of the $Pr^{3+}$ $4f$ electrons, $C_{4f}$, by subtracting the heat capacity of nonmagnetic $La_4Ni_3O_{10}$ from the total $C_p$ of $Pr_4Ni_3O_{10}$. The results are shown in Fig. S19b. Based on the calculated CEF level spectrum, Schottky anomalies associated with the first excited state would be expected to be peaked at $0.42 \cdot \Delta E \approx 20 - 28$ K, in reasonable agreement with the broad feature seen in the magnetic heat capacity centered around 35 K. We find another peak centered at $T_1 = 5$ K. Should this 5 K anomaly arise from a crystal field level poorly calculated by the point charge model, its energy would lie $\approx$12 K (1 meV) above the ground state. More sophisticated models, constrained by inelastic neutron scattering for instance, would be helpful to better understand the crystal field levels and magnetism in $Pr_4Ni_3O_{10}$. The 5 K feature has also been reported by Rout *et al.* [8] who attribute it to a magnetic phase transition that emerges from a notional doublet ground state at Pr A. We note again the low point symmetry, which should not support such a degeneracy. Another possibility is that at low temperature, the product of the induced moment and the internal exchange field at Pr A is sufficiently large with respect to the first excited state CEF level to drive a thermodynamic phase transition, reflected in $C_{4f}$ [14].

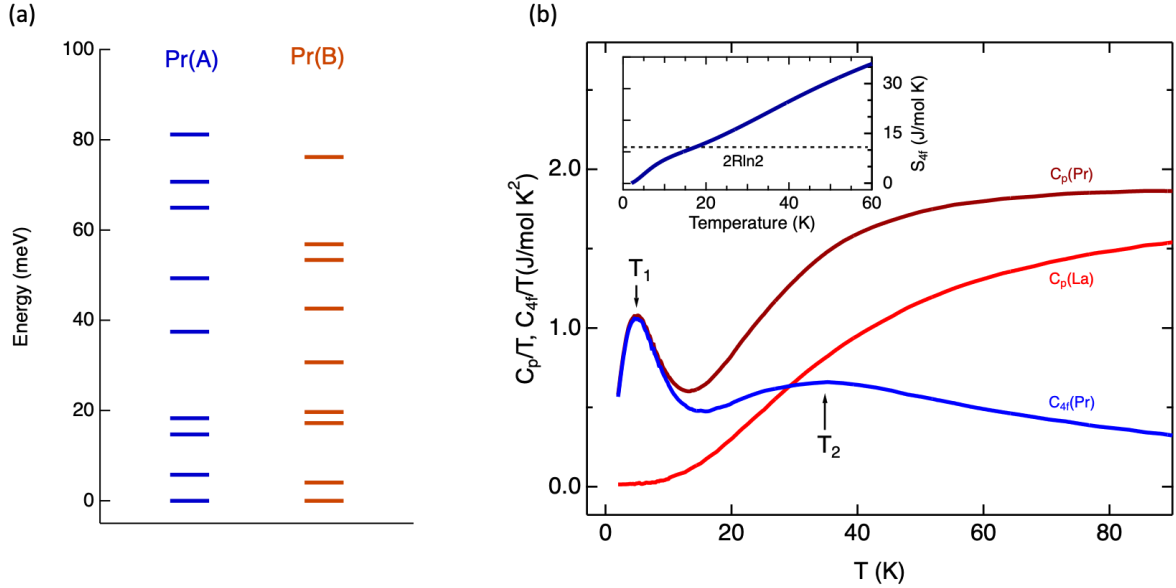

Figure S19: (a) Crystal field levels of $Pr^{3+}$ in $Pr_4Ni_3O_{10}$ calculated using a point charge model. (b) Magnetic heat capacity of the $Pr^{3+}$ $4f$ electrons, $C_{4f}$, obtained by subtracting the Ni and lattice contributions using nonmagnetic $La_4Ni_3O_{10}$ as a background. Inset: magnetic entropy. The data in (b) agree well with those of Fig. 7b of Rout *et al.* [8].